\documentstyle[preprint,aps,psfig,epsfig]{revtex}
\tighten

\begin{document} 
\def\eqn#1{Eq.$\,$#1}
\def\mb#1{\setbox0=\hbox{$#1$}\kern-.025em\copy0\kern-\wd0
\kern-0.05em\copy0\kern-\wd0\kern-.025em\raise.0233em\box0}
\draft
\preprint{}

\title{The statistics of velocity fluctuations arising from a random
distribution of point vortices: the speed of fluctuations and the diffusion
coefficient}
\author{Pierre-Henri Chavanis and Cl\'ement Sire}
\address{Laboratoire de Physique Quantique (UMR C5626 du CNRS), Universit\'e
Paul Sabatier\\ 
118 route de Narbonne, 31062 Toulouse Cedex 4, France\\
(chavanis{@}irsamc2.ups-tlse.fr \& clement{@}irsamc2.ups-tlse.fr)} 
\date{Version of January 18, 2000}
\maketitle 

\begin{abstract}
This paper is devoted to a statistical analysis of the fluctuations of velocity
and acceleration produced by a random distribution of point vortices in
two-dimensional turbulence. We show that the velocity probability density
function (p.d.f.) behaves in a manner  which is intermediate between Gaussian and
L\'evy laws while the distribution of accelerations is governed by a Cauchy law.
Our study accounts properly for a spectrum of circulations among the vortices.
In the case of real vortices (with a finite core) we show analytically that the
distribution of accelerations makes a smooth transition from Cauchy (for small
fluctuations) to Gaussian (for large fluctuations) passing probably  through an
exponential tail.   We introduce a function $T(V)$ which gives the typical
duration of a velocity fluctuation $V$; we show that $T(V)$ behaves like  $V$
and $V^{-1}$ for weak and large velocities respectively. These results have a
simple physical interpretation in the nearest neighbor approximation and in
Smoluchowski (1916) theory concerning the persistence of fluctuations. We
discuss the analogies with respect to the fluctuations of the gravitational
field in stellar systems. As an application of these results, we determine an
approximate expression for the diffusion coefficient of point vortices. When
applied to the context of freely decaying  two-dimensional turbulence, the
diffusion becomes  anomalous and we establish a relationship $\nu=1+{\xi\over
2}$ between the exponent of anomalous diffusion $\nu$ and the exponent $\xi$
which characterizes the decay of the vortex density. This relation is in good
agreement with laboratory experiments and numerical simulations.
\end{abstract}


\newpage

\section{Introduction}
\label{sec_introduction}

A basic problem in fluid turbulence is the characterization  of the entire
stochastic variation of the velocity field ${\bf V}({\bf r},t)$ produced by the
disordered motion of the flow. The velocity   fluctuations can be described by
different quantities such as their  probability density function (p.d.f.), their
typical duration and their  spatial or temporal correlations. We consider a
simple model of  two-dimensional turbulence for which it is possible to
calculate these  quantities exactly. In our model, the velocity is produced by a
collection of point vortices randomly distributed in the domain with uniform
probability. Point vortices behave like particles in interaction and share some
common features with electric charges or stars \footnote{The analogy between
two-dimensional vortices and stellar systems is discussed in detail in
\cite{chav96,chav98b} Chavanis (1996,1998b) and \cite{csr96} Chavanis {\it et
al.} (1996).}. In particular, the problem at hand is directly connected with the
problem of the fluctuations of the  electric field in a gas composed of simple
ions or the fluctuations of the gravitational field  produced by a random
distribution of stars.  These problems were considered  by \cite{holtsmark}
Holtsmark (1919) in electrostatics  and by \cite{chandr} Chandrasekhar (1941)
and \cite{cn42,cn43} Chandrasekhar  \& von Neumann (1942,1943) in a stellar
context. We will show that many  of the methods introduced by these authors can
be extended  to the case of point vortices, even if the calculations and the
results differ due to the lower dimensionality of space ($D=2$ instead of $D=3$)
and the different nature of the interactions.

We consider a collection of $N$ point vortices randomly distributed in a disk of
radius $R$. We assume that the vortices have a Poisson distribution, i.e. their
positions are independent  and  uniformly distributed over the entire domain. We
are particularly interested in the ``thermodynamical limit'' in which the number
of vortices and the  size of the domain go to infinity ($N\rightarrow\infty$,
$R\rightarrow\infty$)  in such a way that the vortex density $n={N\over \pi
R^{2}}$ remains finite. In this limit, the Poisson distribution is shown to be
stationary (see, e.g., \cite{novikov} Novikov 1975) and is well-suited to the
analysis of the fluctuations. For the moment, the vortices have the same
circulation $\gamma$, but we shall indicate later  how the results can be
generalized to include a spectrum of circulations.

The velocity ${\bf V}$ occurring at a given location of the flow is the 
sum of the velocities ${\bf \Phi}_{i}$ $(i=1,...,N)$ produced by
the $N$ vortices:
\begin{equation}
{\bf V}=\sum_{i=1}^{N}{\bf \Phi}_{i}
\label{vtot}
\end{equation}
\begin{equation}
{\bf \Phi}_{i}=-{\gamma\over 2\pi}{{\bf r}_{ \perp i} \over r_{i}^{2}}  
\label{vi}
\end{equation}
where ${\bf r}_{i}$ denotes the position of the $i^{th}$ vortex 
relative to the point under consideration and, by 
definition,  ${\bf r}_{\perp i}$ is the vector ${\bf r}_i$ rotated by 
$+{\pi\over 2}$. Since the vortices are randomly distributed, the velocity 
${\bf V}$ fluctuates. It is therefore of interest to study the statistics 
of these fluctuations, i.e. the probability $W({\bf V})d^{2}{\bf V}$ that 
${\bf V}$ lies between ${\bf V}$ and ${\bf V}+d{\bf V}$. We find that
this distribution behaves in a manner which is
intermediate between Gaussian and L\'evy laws: the core of the distribution
is Gaussian with ``variance'' 
\begin{equation}
\langle V^{2}\rangle ={n\gamma^{2}\over 4\pi}\ln N
\label{varintro}
\end{equation}
while the high velocity  tail decreases algebraically like $V^{-4}$. Since the
``variance'' behaves like $\sim \ln N$, the thermodynamical limit is not   well
defined and the results  are polluted by logarithmic corrections. Previous
investigations of this problem were carried out numerically  by \cite{min} Min
{\it et al.} (1996) and \cite{weiss} Weiss {\it et al.} (1998) and analytically
by \cite{jimenez} Jim\'enez (1996).

However, we must be aware that the knowledge of $W({\bf V})$ alone  does not
provide us  with all the necessary information concerning the fluctuations of
${\bf V}$. An important aspect of the problem concerns the  {\it speed of
fluctuations}, i.e. the typical duration $T({ V})$ of  the velocity fluctuation
${\bf V}$. This requires the knowledge of  the bivariate probability $W({\bf
V},{\bf A}) d^{2}{\bf V}d^{2}{\bf A}$  to measure simultaneously  a velocity
${\bf V}$ with a rate of change
\begin{equation}
{\bf A}={d{\bf V}\over dt}=\sum_{i=1}^{N}{\mb\psi}_{i}
\label{A}
\end{equation}
\begin{equation}
{\mb{\psi}}_{i}=-{\gamma\over 2\pi}\biggl ( {{\bf v}_{\perp i}\over
r_{i}^{2}}-{2({\bf r}_{i}\cdot {\bf v}_{i}){\bf r}_{\perp i}\over
r_{i}^{4}}\biggr )
\label{psi}
\end{equation}
where ${\bf v}_{i}={d{\bf r}_{i}\over dt}$ is the velocity of vortex $i$. Then,
the duration $T({ V})$ can  be estimated by the formula:
\begin{equation}
T({V})={|{\bf V}|\over\sqrt{\langle A^{2}\rangle_{\bf{V}}}}
\label{TV}
\end{equation}
where
\begin{equation}
\langle {A^{2}}\rangle_{\bf V}={\int W({\bf V},{\bf A})A^{2}d^{2}{\bf A}\over
W({\bf V})}
\label{varA}
\end{equation}
is the mean square acceleration associated with a velocity fluctuation   ${\bf
V}$. A similar quantity was introduced by \cite{cn42,cn43} Chandrasekhar  \& von
Neumann (1942,1943) in a stellar context.  We find that the distribution of the
accelerations is governed by a Cauchy law and that the typical duration $T(V)$
of a velocity  fluctuation ${\bf V}$ behaves like $V$ for  $V\rightarrow 0 $ and
$V^{-1}$ for $V\rightarrow\infty$. We also establish that the average duration
of the velocity fluctuations is
\begin{equation}
\langle T\rangle\sim {1\over n\gamma\sqrt{\ln N}}
\label{Taverageint}
\end{equation}
These results can be understood in  the ``nearest neighbor approximation'' in
which the most proximate vortex  plays a prevalent role. In this approximation,
the determination of the  speed of fluctuations can be deduced from the theory
of Smoluchowski (1916)  concerning the mean lifetime of a state with $X$
particles.

In terms of the previous quantities, we can estimate the diffusion coefficient
of point vortices by the formula:
\begin{equation}
D={1\over 4}\int T({\bf V})W({\bf V})V^{2}d^{2}{\bf V}
\label{diff}
\end{equation} 
We find that
\begin{equation}
D\sim \gamma\sqrt{\ln N}
\label{diffexp}
\end{equation}
and we discuss qualitatively how the formation of
``pairs'' modifies the results of our study. In the context of freely decaying
two-dimensional turbulence, the diffusion coefficient is time-dependent (since
the circulation of a vortex increases as a result of successive mergings) and
the diffusion is  anomalous. From equation (\ref{diffexp}), we establish a
relationship 
\begin{equation}
\nu=1+{\xi\over 2}
\label{nuxiint}
\end{equation}
between the exponent of anomalous diffusion $\nu$ and the exponent $\xi$  which
characterizes the decay of the vortex density. This relation is in good
agreement with laboratory experiments and numerical simulations.

We indicate how our results are modified when we allow for a spectrum of
circulations among the vortices. This is an important generalization since
decaying 2D turbulence possesses a continuous distribution of vortices. We show
that the distribution of velocity and acceleration (or velocity gradients) are
only slightly modified by the polydispersity of the vortices and we justify the
validity of previous comparisons of full numerical simulations with vortex
models that ignored this difference (e.g, \cite{jimenez} Jim\'enez 1996).

Finally, we generalize our results to the case of vortex ``blobs'' with a finite
core. We show that the natural cut-off at $r=a$, the vortex radius, removes the
algebraic tail of the velocity distribution. Further, we show analytically that
the distribution of accelerations makes a smooth transition from Cauchy (for
small fluctuations) to Gaussian (for large fluctuations). It is likely that in
between the distribution passes through an {\it exponential tail} as observed
numerically by \cite{min} Min {\it et al.} (1996).

\section{The statistics of velocity fluctuations}
\label{sec_statvelocity}

\subsection{The general expression for $W({\bf V})$}
\label{sec_WV}

We shall now obtain the distribution $W_{N}({\bf V})$ of the velocity ${\bf V}$
produced by $N$ point vortices randomly distributed in a disk of radius $R$
with uniform probability. To avoid a possible solid rotation, we shall assume
that the system is ``neutral'', the circulation of the vortices taking only two
values $+\gamma$ and $-\gamma$ in equal proportion (the case of a non neutral
system is treated in Appendix \ref{sec_nonneutral}). Since a vortex with
circulation $-\gamma$ located in ${\bf r}$ produces the same velocity as a
vortex with circulation $+\gamma$ located in $-{\bf r}$, and since the vortices
are randomly distributed over the entire domain with uniform probability, the
group of vortices with negative circulation is statistically equivalent to the
group of vortices with positive circulation. We can therefore proceed as if
there were a single species of particles but no solid rotation. Since we shall
ultimately let $R\rightarrow \infty$, we can assume without loss of generality
that ${\bf V}$ is calculated at the center of the domain.
    
Under these circumstances, the  distribution $W_{N}({\bf V})$ can be expressed
as:
\begin{equation}
W_{N}({\bf  V})=\int \prod_{i=1}^{N}\tau({\bf r}_{i})d^{2}{\bf
r}_{i}\delta\biggl ({\bf  V}-\sum_{i=1}^{N}{\bf \Phi}_{i}\biggr )
\label{W1}
\end{equation}
where $\tau({\bf r}_{i})d^{2}{\bf r}_{i}$ governs the probability of occurrence
of the $i$-th point vortex at position ${\bf r}_{i}$. In writing this
expression, we have assumed that the vortices are identical and uncorrelated.
Now, using a method originally due to Markov, we express the $\delta$-function
appearing in (\ref{W1}) in terms of its Fourier transform
\begin{equation}
\delta({\bf x})={1\over (2\pi)^{2}}\int e^{-i {\mb \rho}{\bf x}}d^{2}{\mb \rho}
\label{delta}
\end{equation}
With this transformation, $W_{N}({\bf V})$ becomes
\begin{equation}
W_{N}({\bf V})={1\over 4\pi^{2}}\int A_{N}({\mb \rho})e^{-i {\mb \rho}{\bf
V}}d^{2}{\mb\rho} 
\label{W2}
\end{equation}
with:
\begin{equation} 
A_{N}({\mb \rho})=\biggl (\int_{|{\bf r}|=0}^{R}e^{i {\mb \rho}{\bf \Phi}}\tau
({\bf r})d^{2}{\bf r}\biggr )^{N}
\label{AN1}
\end{equation}
where we have written
\begin{equation}
{\bf\Phi}=-{\gamma\over 2\pi}{{\bf r}_{\perp}\over r^{2}}
\label{Phir}
\end{equation}
If we now suppose that the vortices are uniformly distributed  on average, then
\begin{equation}
\tau({\bf r})={1\over \pi R^{2}}
\label{tau}
\end{equation}
and equation (\ref{AN1}) reduces to 
\begin{equation}
A_{N}({\mb \rho})=\Biggl ({1\over \pi R^{2}}\int_{|{\bf r}|=0}^{R} 
e^{i {\mb \rho}{\bf \Phi}}d^{2}{\bf r}\Biggr )^{N}
\label{AN2}
\end{equation}
Since
\begin{equation}
{1\over \pi R^{2}}\int_{|{\bf r}|=0}^{R}d^{2}{\bf r}=1 
\label{norm1}
\end{equation}
we can rewrite our expression for $A_{N}({\mb \rho})$ in the form  
\begin{equation}
A_{N}({\mb \rho})=\Biggl (1-{1\over \pi R^{2}}\int_{|{\bf r}|=0}^{R} 
(1-e^{i {\mb \rho}{\bf \Phi}})d^{2}{\bf r}\Biggr )^{N}
\label{AN3}
\end{equation}
We now consider the limit when the number of vortices and the size of the domain
go to infinity in such a way that the density remains finite: 
$$\qquad
N\rightarrow\infty,\quad  R\rightarrow \infty, \quad  n={N\over \pi R^{2}} \quad
{\rm finite} 
$$ 
If the integral occurring in equation (\ref{AN3}) increases less rapidly than
$N$, then 
\begin{equation}
A({\mb \rho})= e^{-n C({\mb \rho})}
\label{A1}
\end{equation} 
with
\begin{equation}
C({\mb \rho})=\int_{|{\bf r}|=0}^{R} 
(1-e^{i {\mb \rho}{\bf \Phi}})d^{2}{\bf r}
\label{C1}
\end{equation} 
We have dropped the subscript $N$ to indicate that the limit
$N\rightarrow\infty$, in the previous sense,  has been taken. Note that $A({\mb
\rho})$ can still depend on $N$ through logarithmic factors, so that  equation
(\ref{A1}) must be considered as an equivalent of (\ref{AN3}) for large $N$'s,
not a true limit.

To calculate $C({\mb \rho})$ explicitly, it is more convenient to introduce
${\bf \Phi}$ as a variable of integration instead of ${\bf r}$. The Jacobian of
the transformation $\lbrace{\bf r}\rbrace\rightarrow \lbrace{\bf \Phi}\rbrace$
is
\begin{equation}
\biggl |\biggl | {\partial ({\bf r})\over\partial ({\bf \Phi})}\biggr |\biggr
|={\gamma^{2}\over 4\pi^{2}\Phi^{4}}
\label{jacobian1}
\end{equation} 
so that
\begin{equation}
C({\mb \rho})={\gamma^{2}\over 4\pi^{2}}\int_{|{\bf \Phi}|={\gamma\over 2\pi
R}}^{+\infty}(1-  e^{i {\mb \rho}{\bf \Phi}}){1\over \Phi^{4}}d^{2}{\bf \Phi}
\label{C2}
\end{equation}
or, alternatively,
\begin{equation}
C({\mb \rho})={\gamma^{2}\over 4\pi^{2}}\int_{|{\bf \Phi}|={\gamma\over 2\pi
R}}^{+\infty}
(1-\cos ( {\mb \rho}{\bf \Phi})){1\over \Phi^{4}}d^{2}{\bf \Phi}
\label{C3}
\end{equation} 
Chosing polar coordinates with the $x$-axis in the direction of $\mb {\rho}$,
equation (\ref{C3}) can be transformed to
\begin{equation}
C({\mb \rho})={\gamma^{2}\over 4\pi^{2}}\int_{\gamma\over 2 \pi
R}^{+\infty}{d\Phi\over\Phi^{3}}\int_{0}^{2\pi} (1-\cos( \rho 
\Phi\cos\theta))d\theta
\label{C4}
\end{equation}  
where $\theta$ denotes the angle between ${\mb \rho}$ and  ${\bf \Phi}$. Using
the identity
\begin{equation}
\int_{0}^{\pi}\cos (z\cos \theta)d\theta=\pi J_{0}(z)
\label{identity1}
\end{equation}  
we obtain
\begin{equation}
C({\mb \rho})={\gamma^{2}\over 2\pi}\int_{\gamma\over 2 \pi
R}^{+\infty}(1-J_{0}(\rho\Phi)){d\Phi\over \Phi^{3}}
\label{C5}
\end{equation}  
or, writing $x=\rho\Phi$,
\begin{equation}
C({\mb \rho})={\gamma^{2}\rho^{2}\over 2\pi}\int_{\gamma\rho\over 2 \pi
R}^{+\infty} (1-J_{0}(x)){dx\over x^{3}}
\label{C6}
\end{equation}
Recall that $C({\mb\rho})$ must be evaluated in the limit
$N,R\rightarrow\infty$ with  $n={N\over \pi R^{2}}$ finite. Using the well-known
expansion of the Bessel function $J_{0}$ for small arguments:
\begin{equation}
J_{0}(x)=1-{x^{2}\over 4}+o(x^{4})
\label{expJ0}
\end{equation}  
we have the estimate:
\begin{equation}
C({\mb \rho})={\gamma^{2}\rho^{2}\over 16\pi}\ln \biggl ( {4\pi N\over
n\gamma^{2}\rho^{2}}\biggr )
\label{Crhogeneral}
\end{equation}
Since $C({\mb \rho})$ diverges weakly with $N$ (logarithmically), the limiting
process leading to formula (\ref{A1}) is permissible. For $\rho>0$ and
$N\rightarrow \infty$, we have
\begin{equation}
A({\mb \rho})=e^{-{n\gamma^{2}\over 16\pi}\ln N \rho^{2} }
\label{Alarge}
\end{equation}
and for $\rho\rightarrow 0$, we obtain
\begin{equation}
A({\mb \rho})=e^{{n\gamma^{2}\over 8\pi}(\ln \rho) \rho^{2} }
\label{Asmall}
\end{equation}
The velocity distribution  $W({\bf  V})$ is simply the Fourier transform of
$A({\mb\rho})$. We shall now derive the expression for $W({\bf V})$ in the core
and in the tail of the distribution.

\subsection{The core of the distribution $W({\bf V})$}
\label{sec_core}

For $V\lesssim V_{crit}(N)$, where $V_{crit}(N)$ is defined by formula
(\ref{Vcrit}), the contribution of small $\rho$'s in the integral (\ref{W2}) is
negligible and we can use the expression (\ref{Alarge}) for $A({\mb\rho})$. In
that case, the distribution $W({\bf  V})$ is the Gaussian
\begin{equation}
W({\bf  V})= {4\over  {n\gamma^{2}} \ln N} e^{-{4\pi\over n {\gamma^{2}}\ln N}
{V}^{2}}\quad (V\lesssim V_{crit}(N))
\label{Wgauss}
\end{equation} 
If we were to extend this distribution for all values of $V$, we would conclude
that its variance
\begin{equation}
\langle V^{2}\rangle = {n {\gamma^{2}} \over 4\pi} \ln N 
\label{variance}
\end{equation}
diverges logarithmically when $N\rightarrow \infty$. This result was noted  by
\cite{jimenez} Jim\'enez (1996), \cite{min} Min {\it et al.} (1996) and
\cite{weiss} Weiss {\it et al.} (1998) who  applied a generalized form of the
Central Limit Theorem.  In fact, the Central Limit Theorem is not strictly
applicable here because the variance of the velocity created by a single vortex
\begin{equation} 
\langle \Phi^{2}\rangle=\int_{|{\bf r}|=0}^{R}\tau({\bf
r})\Phi^{2}d^{2}{\bf r}=\int_{0}^{R}{1\over\pi R^{2}}{\gamma^{2}\over
4\pi^{2}r^{2}}2\pi r dr
\label{vari}
\end{equation}
diverges logarithmically when $r\rightarrow 0$; still, the distribution of ${\bf
V}$ is Gaussian (for $V\lesssim V_{crit}(N)$) but its ``variance'' behaves like
$\ln N$. For $V\gtrsim V_{crit}(N)$, the distribution (\ref{Wgauss}) breaks down
because, for large velocities, the Fourier transform (\ref{W2}) is dominated by
the contribution of small $\rho$'s and formula (\ref{Asmall}) must be used
instead of (\ref{Alarge}).  This implies that the high velocity tail of the
distribution $W({\bf V})$ decays algebraically  like $V^{-4}$ (see section
\ref{sec_tail}). This algebraic tail  arises because we are on the frontier
between Gaussian and L\'evy laws (see figure 1.1 of \cite{bouchaud} Bouchaud \&
Georges 1990 and section \ref{sec_NNA}).

The distribution (\ref{Wgauss}) has been derived for a neutral system consisting
in an equal number of vortices with circulation $+\gamma$ and $-\gamma$. If the
system is non-neutral, there is a solid rotation and the average velocity
increases linearly with the distance. Therefore, at point ${\bf a}$, equation
(\ref{Wgauss}) must be replaced by (see Appendix \ref{sec_nonneutral}):
\begin{equation} 
W({\bf V})= {4\over  {n\gamma^{2}} \ln N} e^{-{4\pi\over n {\gamma^{2}}\ln N}
{({\bf V}-{1\over 2}n\gamma {\bf a}_{\perp})^{2}}}\quad (|{\bf V}-{1\over
2}n\gamma {\bf a}_{\perp}|\lesssim V_{crit}(N))
\label{Wgn}
\end{equation}
The velocity distribution at ${\bf a}\neq {\bf 0}$ differs only from the
distribution at the center of the domain by replacing the velocity ${\bf V}$ by
the fluctuating velocity ${\mb {\cal V}}={\bf V}-\langle {\bf V}\rangle={\bf
V}-{1\over 2}n\gamma {\bf a}_{\perp}$. A factor $1/2$ arises in front of the
average vorticity $n\gamma$ because, for a solid rotation, the vorticity is
twice the angular velocity. In Appendix \ref{sec_generalization}, we extend our
results to allow for a spectrum of circulations among the vortices.

\subsection{The high velocity tail of the distribution $W({\bf V})$}
\label{sec_tail}

We shall now determine the behavior of the distribution $W({\bf V})$ for
$V\rightarrow\infty$. Introducing polar coordinates with the $x$-axis in the
direction of ${\bf V}$, and using (\ref{A1}), equation (\ref{W2}) can be
transformed to
\begin{equation}
W({\bf V})={1\over 2\pi^{2}}\int_{0}^{\pi}d\theta\int_{0}^{+\infty} 
e^{-i {\rho}{V}\cos\theta}e^{-nC({ \rho})}\rho d\rho
\label{Wt1}
\end{equation}
With the change of variables $z=\rho V$ 
and $t=-\cos\theta$, equation (\ref{Wt1}) can be rewritten
\begin{equation}
W({\bf V})={1\over 2\pi^{2}V^{2}}{\cal R}_{e}\int_{-1}^{+1}{dt\over 
\sqrt{1-t^{2}}}\int_{0}^{+\infty} e^{i z t}e^{-nC({z\over V})}z dz
\label{Wt2}
\end{equation} 
In this expression, $t$ and $z$ are real and the domains of integration
$\tau_{0}: -1\le t \le 1$ and $\zeta_{0}: 0\le z<+\infty$ are taken along  the
real axis. Under these circumstances, the integral is not convergent  if we
expand the quantity  $$e^{-nC({z\over V})}$$ in a power series of ${z\over V}$,
for $V\rightarrow +\infty$,  and evaluate the integral term by term. However,
regarding $z$ and $t$ as complex variables, it is possible to choose paths of
integration along  which this expansion will converge \footnote{This method is
inspired from \cite{cn42} Chandrasekhar  \& von Neumann (1942).}.

We shall first carry out the integration on $z$, for a fixed $t$. It will
therefore be possible to choose the (complex) integration paths for
$z$ dependent on $t$. The integration paths are
modified as follows: $\tau_{0}$ is replaced by $\tau$, the semi-circle 
with radius unity lying in the domain ${\cal I}_{m}(t)\ge 0$. Therefore, 
$\arg(t)$ varies from $\pi$ to $0$ when $t$ moves from $-1$ to 
$+1$. On the other hand, $\zeta_{0}$ is replaced by $\zeta_{\omega_{t}}$, the
line starting from the origin and forming an angle 
\begin{equation}
\omega_{t}={1\over 8}\biggl ({\pi\over 2}-\arg (t)\biggr)
\label{angle1}
\end{equation}      
with the real axis. When $t$ moves from $-1$ to $+1$, $\omega_{t}$ varies  from
$-{\pi\over 16}$ to $+{\pi\over 16}$. For $|z|\rightarrow\infty$, according to
(\ref{Alarge}), we have:
\begin{equation}
e^{-nC({z\over V})}=e^{-{n\gamma^{2}\over 16\pi}\ln N {z^{2}\over V^{2}}}
\label{enC1}
\end{equation}   
Since the argument of $z^{2}$ is between $-{\pi\over 8}$ and ${\pi\over 8}$, its
real part is always positive and the convergence of equation (\ref{Wt2}) is
undisturbed. On the other hand, the argument of $i z t$ is equal to $${\pi\over
2}+{1\over 8}\biggl ({\pi\over 2}-\arg (t)\biggr)+\arg(t)$$     and lies between
${9\pi\over 16}$ and ${23\pi\over 16}$.  Therefore, the real part of $i z t$  is
always negative and the function $e^{i z t}$    decays exponentially to zero as
$|z|\rightarrow\infty$. Therefore, with the new paths  of integration $\tau$ and
$\zeta_{\omega_{t}}$, it is possible to expand the integrand of equation
(\ref{Wt2})  in power series of ${z\over V}$,  for $V\rightarrow\infty$, and
integrate term by term. When  ${z\over V}\rightarrow 0$, we have, according to
(\ref{Asmall}), 
\begin{equation}
e^{-n C ({z\over V})}=e^{{n\gamma^{2}\over 8\pi}\ln ({z\over
V}) {z^{2}\over V^{2}}} \label{enC2} 
\end{equation}  
and we can write:
\begin{equation}
W({\bf V})={1\over 2\pi^{2}V^{2}}{\cal R}_{e}\int_{\tau}{dt\over 
\sqrt{1-t^{2}}}\int_{\zeta_{\omega_{t}}} e^{i z t}\biggl\lbrack 1+{n 
\gamma^{2}\over 8\pi}\ln\biggl ({z\over V }\biggr ) {z^{2}\over V^{2}}+...
\biggr\rbrack  z dz
\label{Wt3} 
\end{equation} 
Since this integral is convergent along any line on which the real  part of $i z
t$ is negative, we can replace the integration path  $\zeta_{\omega_{t}}$ by the
line $\zeta_{\psi_{t}}$ forming an angle 
\begin{equation}
\psi_{t}={\pi\over 2}-\arg(t)
\label{angle2}
\end{equation}  
with the real axis. On this new integration path
$$ i z t =-y , \qquad y \quad {\rm real}\ge 0$$
and we get
\begin{eqnarray}
W({\bf V})=-{1\over 2\pi^{2}V^{2}}{\cal R}_{e}\int_{-1}^{+1}{dt\over
\sqrt{1-t^{2}}}\int_{0}^{+\infty} e^{-y}\times\nonumber\\
\biggl\lbrack 1-{n \gamma^{2}\over
8\pi}\ln\biggl ({ i y\over V}\biggr ){1\over t^{2}} {y^{2}\over
V^{2}} +{n\gamma^{2}\over 8\pi} {\ln t\over t^{2}}{y^{2}\over
V^{2}}+...\biggr\rbrack  {y\over t^{2}} dy
\label{Wt4}
\end{eqnarray} 
where we recall that $t$ is a complex variable and the integration has 
to be performed over the semi-circle of radius unity lying on the domain
${\cal I}_{m}(t)\ge 0$. Writing $t=e^{i\theta}$, we find that
\begin{equation}
\int_{-1}^{+1}{dt\over t^{2}\sqrt{1-t^{2}}}=0 ;\\ \int_{-1}^{+1}{dt\over
t^{4}\sqrt{1-t^{2}}}=0 ;\\ \int_{-1}^{+1}{\ln t\over
t^{4}\sqrt{1-t^{2}}}dt=-{2\pi\over 3} 
\label{intcomplex1}
\end{equation}  
Therefore, 
\begin{equation}
W({\bf V})={n\gamma^{2}\over 24\pi^{2}V^{4}}\int_{0}^{+\infty} e^{-y} y^{3} dy
\label{Wt5}
\end{equation} 
In this expression, we recognize the $\Gamma$-function
\begin{equation}
\Gamma(n+1)= \int_{0}^{+\infty} e^{-y} {y}^{n} dy
\label{Gamma}
\end{equation}
with $n=3$. Its value is $\Gamma(4)=6$ and  we finally obtain
\begin{equation}
W({\bf V})= {n\gamma^{2}\over 4 \pi^{2}V^{4}}\quad (V\gtrsim V_{crit}(N))
\label{Wt6}
\end{equation}
Therefore, for sufficiently large values of $V$, the velocity  distribution
$W({\bf V})$ decays algebraically, like $V^{-4}$. In section \ref{sec_NNA}, we
give a  physical interpretation of this result in terms of the nearest neighbor
approximation. 

>From equations (\ref{Wgauss}) and (\ref{Wt6}), we can estimate the value of the
velocity $V_{crit}(N)$ for which the distribution $W({\bf V})$ departs from the
Gaussian. This $V_{crit}(N)$ is obtained by seeking  the point where the two
regimes (\ref{Wgauss}) and (\ref{Wt6})   connect each other. Neglecting
subdominant terms in $\ln N$, one finds simply
\begin{equation}
V_{crit}(N)\sim \biggl ({n\gamma^{2}\over 4\pi}\ln N\biggr )^{1/2}\ln^{1/2}(\ln
N)
\label{Vcrit}
\end{equation}
This result shows that the convergence to a pure Gaussian distribution  is
extremely slow with $N$ as emphasized by \cite{jimenez} Jim\'enez (1996),
\cite{min} Min {\it et al.} (1996) and \cite{weiss} Weiss {\it et al.} (1998).
Since the distribution $W({\bf V})$ decreases like $V^{-4}$ for
$V\rightarrow\infty$, the variance of the velocity diverges logarithmically.

Note, finally, that the distribution of $V_{x}$, the $x$-component of the
velocity is
\begin{equation}
W(V_{x})={2\over\sqrt{n\gamma^{2}\ln N}}e^{-{4\pi V_{x}^{2}\over n\gamma^{2}\ln
N}}\quad (V\lesssim  V_{crit}(N))
\label{WVx}
\end{equation}
\begin{equation}
W(V_{x})={n\gamma^{2}\over 8\pi V_{x}^{3}}\quad (V\gtrsim V_{crit}(N))
\label{WVy}
\end{equation}

\subsection{The formation of ``pairs''}
\label{sec_pv}

The previous results should be all the more valid if the velocity ${\bf V}$ is
calculated at a fixed point of the domain. In such a case, there is no
restriction on the possible values of $V$ since a vortex can approach this point
with no limit producing extremely large velocities. The situation is different
if ${\bf V}$ is now the velocity experienced by a ``test'' vortex. Indeed, if a
``field'' vortex approaches the test vortex below a certain distance, then a
``pair'' will form and our treatment, which ignores the correlations between
vortices, will clearly break down. These pairs have been observed and studied
numerically by \cite{weiss} Weiss {\it et al.} (1998). 

We can simply estimate the typical separation below which a pair will form by
comparing the velocity produced by a single vortex ${\gamma\over 2\pi r}$ with
the typical velocity $V_{typ}=({n\gamma^{2}\over 4\pi}\ln N)^{1/2}$ produced by
the  field (see equation \ref{variance}).   This yields 
\begin{equation}
d_{crit}(N)=(\pi n\ln N)^{-1/2}
\label{dcrit}
\end{equation}
a distance slightly smaller than  the interparticle distance by a factor $\sim
1/\sqrt{\ln N}$. In the mathematical limit $N\rightarrow\infty$, there is no
pair, since $d_{crit}\rightarrow 0$. This result is in agreement with the
stationarity of the Poisson process when $N\rightarrow\infty$: if the vortices
are initially uncorrelated, they will remain uncorrelated. However, the
convergence is extremely slow with $N$ and close pairs will always form in
realistic situations. As emphasized by  \cite{weiss} Weiss {\it et al.} (1998),
a system of $10^3-10^5$ vortices has a behavior which is a combination of both
low-dimensional behavior, i.e. closed pairs, and high dimensional behavior
described by traditional stochastic processes.

A pair can be either a ``binary''  (rapidly  rotating around its center of
vorticity) when two vortices of the  same sign are bound together, or a
``dipole'' (translating or rotating), when  two vortices of opposite sign pair
off. Of course, binaries and dipoles behave very differently.  If the vortex is
engaged in a binary with long correlation time, then, for  practical purposes,
the relevant velocity to consider is not its own velocity  (which has a rotating
component), but rather the velocity of the center of vorticity which is induced
by the rest of the system. Therefore, a binary simply behaves like a single
point vortex with larger circulation and relatively slow velocity (otherwise
this means that the binary is itself engaged in a pair). It may be noted that,
in the case of  real vortices (with a finite core), the formation of binaries is
replaced by merging events. By contrast, a dipole moves by itself  and  behaves
like a kind of particle undergoing fast ballistic motion. Its velocity may be
large but, since it creates a dipolar velocity field, the previous results
cannot be applied directly and an appropriate treatment is required.

We therefore expect that the velocity distribution
(\ref{Wgauss})(\ref{Wt6}) which ignores correlations between vortices will
break down for $V\gg V_{crit}(N)$ since, in that case, the velocity is entirely
due to the nearest neighbor and pairs form. In the following, we shall account
for this failure by introducing a cuf-off at some $V_{max}$, i.e. 
\begin{equation}
W({\bf V})=0 \qquad (V>V_{max})
\label{Wcutoff}
\end{equation}  
This is the simplest modification that we can make to account for the formation
of pairs at large velocities. For vortices with size $a$, the Gaussian
distribution (\ref{Wgauss}) is valid for all $V$'s until the natural cut-off at
$V_{max}\sim {\gamma\over 2\pi a}$ (see Appendix \ref{sec_blobs}).

\section{The statistics of accelerations}
\label{sec_statacceleration}

\subsection{The general formula for $W({\bf V},{\bf A})$ }
\label{sec_WVA}

We are concerned here with the calculation of the bivariate probability
$W_{N}({\bf V},{\bf A})$ to measure simultaneously a velocity ${\bf V}$  with a
rate of change ${\bf A}={d{\bf V}\over dt}$. According to equations (\ref{vtot})
and (\ref{A}), ${\bf V}$ and ${\bf A}$ are the sum  of $N$ random variables
${\bf \Phi}_{i}$ and ${\mb \psi}_{i}$ depending on  the positions ${\bf r}_{i}$
and velocities ${\bf v}_{i}$ of the point  vortices. However, unlike material
particles, the variables $\lbrace {\bf r}_{i}$, ${\bf v}_{i}\rbrace$, for
different $i$'s, are not independent because the velocities of the vortices are
determined by the configuration  $\lbrace {\bf r}_{i}\rbrace$ of the system as a
whole. However, for our purpose, it is probably a reasonable approximation to
neglect these  correlations and treat $\lbrace {\bf r}_{i}$, ${\bf
v}_{i}\rbrace$ ($i=1,...,N$) as independent  variables. We shall only describe
qualitatively how the formation of pairs affects our results.

When this decorrelation hypothesis is implemented, a straightforward
generalization of the method used in section \ref{sec_WV} yields 
\begin{equation}
W_{N}({\bf V},{\bf A})={1\over 16\pi^{4}}\int   e^{-i ({\mb \rho}{\bf V}+{\mb
\sigma}{\bf A})}A_{N}({\mb \rho},{\mb \sigma})d^{2}{\mb \rho}d^{2}{\mb \sigma}
\label{WW1}
\end{equation}    
with 
\begin{equation}
A_{N}({\mb \rho},{\mb \sigma})=\biggl (\int_{|{\bf r}|=0}^{R}\int_{|{\bf
v}|=0}^{+\infty} \tau({\bf r},{\bf v})e^{i({\mb \rho}{\bf \Phi}+{\mb
\sigma}{\mb\psi})}d^{2}{\bf r}d^{2}{\bf v}\biggr )^{N}
\label{AA1}
\end{equation}    
where we have defined
\begin{equation}
{\bf \Phi}=-{\gamma\over 2\pi}{{\bf r}_{\perp}\over r^{2}}
\label{aPhi}
\end{equation}  
\begin{equation}
{\mb \psi}=-{\gamma\over 2\pi}\biggl ( {{\bf v}_{\perp }\over r_{}^{2}}-{2({\bf
r}\cdot {\bf v}){\bf r}_{\perp }\over r^{4}}\biggr )
\label{apsi}
\end{equation}
and where $\tau({\bf r},{\bf v})$ denotes the probability that a vortex be in
${\bf r}$ with velocity ${\bf v}$. According to our initial assumptions, the
vortices are distributed uniformly on average and $\tau({\bf r})$ is given by
equation (\ref{tau}). On the other hand, their velocity distribution $\tau({\bf
v})$ is given by equations (\ref{Wgauss})(\ref{Wt6}) of section
\ref{sec_statvelocity}. However, due to the formation of pairs, this
distribution must be modified at large velocities (see section \ref{sec_pv}).
Instead of introducing a sharp cut-off $\tau({\bf v})=0$ at $v>v_{max}$, we
shall assume for convenience that the Gaussian distribution (\ref{Wgauss}) is
valid for all velocities \footnote{In Appendix \ref{sec_generalization},  we
generalize our results for an arbitrary isotropic distribution of the velocity
$\tau(|{\bf v}|)$.}. Therefore, the probability that a vortex be in ${\bf r}$
with velocity ${\bf v}$ is 
\begin{equation}
\tau({\bf r},{\bf v})={1\over \pi R^{2}}\times {4\over n\gamma^{2} \ln
N}e^{-{4\pi\over  n\gamma^{2}\ln N}v^{2}} 
\label{df}
\end{equation}
It is remarkable that the distribution (\ref{df}) is formally equivalent to the
Maxwell-Bolzmann statistics of material particles at equilibrium. Owing to this
analogy, we can interpret the variance
\begin{equation}
\overline{v^{2}}={n\gamma^{2}\over 4\pi}\ln N
\label{kt}
\end{equation}    
as a kind of kinetic ``temperature''. More generally, the moment of order $p$ of
the velocity is
\begin{equation}
\overline{v^{p}}=\biggl ({n\gamma^{2}\ln N\over 4\pi}\biggr )^{p\over
2}\Gamma\biggl ({p\over 2}+1\biggr)
\label{vp}
\end{equation}
where the $\Gamma$-function is defined by equation (\ref{Gamma}). In particular
\begin{equation}
\overline{v}=\biggl ({n\gamma^{2}\over 16}\ln N\biggr )^{1\over 2}
\label{v1}
\end{equation}

Recall that the distribution (\ref{df}) is valid only for a neutral system made
of an equal number of vortices with circulation $+\gamma$ and $-\gamma$ (if the
system is non neutral we must account for a solid rotation). Since a vortex with
circulation $-\gamma$ located in ${\bf r}$ and moving with velocity ${\bf v}$
produces the same velocity ${\bf V}$ {\it and} acceleration ${\bf A}$  as a
vortex with circulation $+\gamma$ located in $-{\bf r}$ and moving with velocity
${\bf -v}$, and since the vortices are randomly distributed  with uniform
probability and isotropic velocity distribution, the two groups of vortices  are
statistically equivalent. Therefore, as in section \ref{sec_statvelocity}, we
can proceed as if we had a single type of vortex with circulation $\gamma$ and
no solid rotation. In Appendix \ref{sec_generalization}, we extend our results
to allow for a spectrum of circulations among the vortices (still for a neutral
system) and in Appendix \ref{sec_nonneutral} we consider the case of a
non-neutral system.

Substituting the expression (\ref{df}) for $\tau({\bf r},{\bf v})$ in equation
(\ref{AA1}), we obtain
\begin{equation}
A_{N}({\mb \rho},{\mb \sigma})=\biggl ({1\over \pi R^{2}}\int_{|{\bf
r}|={0}}^{R}\int_{|{\bf v}|=0}^{+\infty}{4\over  n{\gamma^{2}} \ln N} e^{-{4\pi
\over n {\gamma^{2}} \ln N}  v^{2}} e^{i({\mb \rho}{\bf \Phi}+{\mb
\sigma}{\mb\psi})}    d^{2}{\bf r}d^{2}{\bf v }\biggr )^{N}
\label{AA2}
\end{equation} 
As in section \ref{sec_WV}, it is more convenient to use ${\bf \Phi}$ and ${\mb
\psi}$ as variables of integration rather than ${\bf r}$ and ${\bf v}$. The
Jacobian of the transformation $\lbrace {\bf r},{\bf v}\rbrace\rightarrow
\lbrace {\bf \Phi},{\mb \psi}\rbrace$ is
\begin{equation}
\biggl |\biggl | {\partial ({\bf r},{\bf v})\over\partial ({\bf \Phi},{\mb
\psi})}\biggr |\biggr |={\gamma^{4}\over 16\pi^{4}\Phi^{8}}
\label{jacobian2}
\end{equation}
We must next express $v=|{\bf v}|$ in terms of our new variables ${\bf \Phi}$
and ${\mb \psi}$. According to equations (\ref{aPhi}) and (\ref{apsi}), we get
\begin{equation}
\Phi={\gamma\over 2\pi r}\quad {\rm and }\quad \psi=
{\gamma  v\over 2\pi r^{2}}
\label{Phiandpsi}
\end{equation}   
Hence
\begin{equation}
v={\gamma\over 2\pi}{\psi\over\Phi^{2}}
\label{vphipsi}
\end{equation}    
Thus, in these new variables, the expression for $A_{N}({\mb\rho},{\mb\sigma})$
becomes
\begin{equation}
A_{N}({\mb \rho},{\mb \sigma})=\biggl ({1\over \pi R^{2}}\int_{|{\mb
\Phi}|={\gamma\over 2\pi R}}^{+\infty}\int_{|{\mb \psi}|=0}^{+\infty}{4\over  n
{\gamma^{2}} \ln N} e^{-{1\over n\pi \ln N}{\psi^{2}\over \Phi^{4}}} e^{i({\mb
\rho}{\bf \Phi}+{\mb \sigma}{\mb\psi})}{\gamma^{4}\over 16\pi^{4}\Phi^{8}}
d^{2}{\bf \Phi}d^{2}{\mb  \psi}\biggr )^{N}
\label{AA3}
\end{equation}  
The integral with respect to ${\mb \psi}$ is Gaussian and is readily evaluated.
We are left with 
\begin{equation}
A_{N}({\mb \rho},{\mb \sigma})=\biggl ({1\over \pi R^{2}}\int_{|{\mb
\Phi}|={\gamma\over 2\pi R}}^{+\infty}    e^{i{\mb \rho}{\bf \Phi}}e^{-{1\over
4}\pi n\ln N\sigma^{2}\Phi^{4}}{\gamma^{2}\over
4\pi^{2}\Phi^{4}}d^{2}{\bf\Phi}\biggr )^{N}
\label{AA4}
\end{equation}    
We verify that 
\begin{equation}
{1\over \pi R^{2}}\int_{|{\mb \Phi}|={\gamma\over 2\pi R}}^{+\infty}
{\gamma^{2}\over 4\pi^{2}\Phi^{4}}d^{2}{\bf\Phi}=1
\label{norm2}
\end{equation}   
Hence, we can rewrite equation (\ref{AA4}) in the form
\begin{equation}
A_{N}({\mb \rho},{\mb \sigma})=\biggl (1-{1\over \pi R^{2}}\int_{|{\mb
\Phi}|={\gamma\over 2\pi R}}^{+\infty}  (1-  e^{i{\mb \rho}{\bf
\Phi}}e^{-{1\over 4}\pi n\ln N\sigma^{2}\Phi^{4}}){\gamma^{2}\over
4\pi^{2}\Phi^{4}}d^{2}{\bf\Phi}\biggr )^{N}
\label{AA5}
\end{equation}   

We shall now consider the limit $N,R\rightarrow\infty$ with $n={N\over \pi
R^{2}}$ finite.  If the integral appearing in equation (\ref{AA5}) increases
less rapidly than $N$, then
\begin{equation}
A({\mb \rho},{\mb \sigma})=e^{-n C({\mb \rho},{\mb \sigma})}
\label{AA6}
\end{equation}
with 
\begin{equation}
C({\mb \rho},{\mb \sigma})= {\gamma^{2}\over  4\pi^{2}}\int_{|{\mb
\Phi}|={\gamma\over 2\pi R}}^{+\infty} (1- e^{i{\mb \rho}{\bf \Phi}} e^{-{1\over
4} \pi n\ln N\sigma^{2}\Phi^{4}}){1\over \Phi^{4}}d^{2}{\bf\Phi}
\label{CC1}
\end{equation} 
As in section \ref{sec_WV}, the function $A({\mb \rho},{\mb \sigma})$ represents
an equivalent of $A_{N}({\mb \rho},{\mb \sigma})$ for large $N$, not a true
limit. It can therefore still depend on $N$ through logarithmic factors.

After introducing polar coordinates and integrating over the angular variable
using (\ref{identity1}), we  arrive at 
\begin{equation}
C({\mb \rho},{\mb \sigma})= {\gamma^{2}\over 2\pi}\int_{\gamma \over 2\pi
R}^{+\infty} (1- J_{0}({\rho}{\Phi}) e^{-{1\over 4} \pi n\ln
N\sigma^{2}\Phi^{4}}){d\Phi\over \Phi^{3}}
\label{CC2}
\end{equation}  
Equations (\ref{WW1})(\ref{AA6}) and (\ref{CC2}) formally solve the problem but
the integrals look difficult to calculate explicitly.  However, if we are only
interested in the moments  of ${\bf A}$ for a given ${\bf V}$ (resp. in the
moments of ${\bf V}$ for a  given ${\bf A}$), we only need  the asymptotic
expansion of  $A({\mb \rho},{\mb \sigma})$ for $|{\mb \sigma}|\rightarrow 0$
(resp.  $|{\mb \rho}|\rightarrow 0$). This problem will be considered in
sections  \ref{sec_A2V}. First, we shall derive the
unconditional distribution of the  acceleration.

\subsection{The Cauchy distribution for ${\bf A}$}
\label{sec_Cauchy}

According to equation (\ref{WW1}), we clearly have
\begin{equation}
W({\bf A})={1\over 16\pi^{4}}\int A({\mb \rho},{\mb \sigma}) e^{-i ({\mb
\rho}{\bf V}+{\mb \sigma}{\bf A})} d^{2}{\mb \rho}d^{2}{\mb \sigma} 
d^{2}{\bf V} 
\label{WA1}
\end{equation}
Using (\ref{delta}), the foregoing expression for $W({\bf A})$ reduces to
\begin{equation}
W({\bf A})={1\over 4\pi^{2}}\int A({\mb\sigma}) e^{-i{\mb\sigma}{\bf A}} 
d^{2}{\mb\sigma}
\label{WA2}
\end{equation}
where we have written $A({\mb\sigma})$ for $A({\bf 0},{\mb\sigma})$. Hence,
according to equations (\ref{AA6}) and (\ref{CC2}), we get 
\begin{equation}
A({\mb \sigma})=e^{-n C({\mb \sigma})}
\label{AA7}
\end{equation}
with
\begin{equation}
C({\mb \sigma})= {\gamma^{2}\over 2\pi}\int_{0}^{+\infty} 
 (1- e^{-{1\over 4}\pi n\ln N\sigma^{2}\Phi^{4}} ){d\Phi\over \Phi^{3}}
\label{CC3}
\end{equation}
Following the usual prescription, we have let $R\rightarrow \infty$  since the
integral is convergent when $\Phi\rightarrow 0$. Integrating by parts, we obtain
\begin{equation}
C({\mb \sigma})= {\gamma^{2}\over 8}\sqrt{n\ln N}\sigma
\label{CC4}
\end{equation}
Hence
\begin{equation}
A({\mb \sigma})=e^{-{1\over 8}\gamma^{2}n^{3/2}\sqrt{\ln N}\sigma}
\label{AA8}
\end{equation} 
The distribution $W({\bf A})$ is simply the Fourier transform of the exponential
function (\ref{AA8}). This is the 2D-Cauchy distribution:
\begin{equation}
W({\bf A})={32\over \pi n^{3}\gamma^{4}\ln N}{1\over\biggl (1+{64 A^{2}\over
n^{3}\gamma^{4}\ln N}\biggr )^{3/2}} 
\label{WA3}
\end{equation}
The Cauchy distribution is a particular L\'evy law. As such, it decays
algebraically  for large $|{\bf A}|$'s, according to
\begin{equation}
W({\bf A})\sim {\gamma^{2} n^{3/2}\sqrt{\ln N}\over 16\pi A^{3}}\qquad
(A\rightarrow\infty)
\label{WA4}
\end{equation}
This result has a clear physical interpretation in the nearest neighbor
approximation (see section \ref{sec_NNA}). Only the moments of order $p<1$ of
the acceleration  exist and we have the general expression
\begin{equation}
\langle A^{p}\rangle=\biggl ({n^{3/2}\gamma^{2}\sqrt{\ln N}\over 8}\biggr
)^{p}{1\over\sqrt{\pi}}\Gamma\biggl ({1-p\over 2}\biggr )\Gamma\biggl (1+{p\over
2}\biggr )
\label{AP}
\end{equation}
Note, finally, that the distribution of $A_{x}$, the $x$-component of the
acceleration, is the ordinary 1D-Cauchy law
\begin{equation} 
W(A_{x})={8\over \pi n^{3/2}\gamma^{2}\sqrt{\ln N}}{1\over 1+{64 A_{x}^{2}\over
n^{3}\gamma^4\ln N}} 
\label{Cauchy1D} 
\end{equation}
Clearly, the distribution of acceleration is related to the distribution of
velocity gradients $\delta{\bf V}$ (see \cite{jimenez} Jim\'enez 1996 and
\cite{min} Min {\it et al.} 1996)

\subsection{The moments $\langle A^{2}\rangle _{\bf V}$ and $\langle
V^{2}\rangle _{\bf A}$ }
\label{sec_A2V}

Let us introduce a Cartesian system of coordinates and denote by $\lbrace
A_{\mu}\rbrace$  the different components of ${\bf A}$ relative to that frame.
The average value of $A_{\mu}A_{\nu}$ for a given velocity ${\bf V}$ is defined
by 
\begin{equation}
\langle {A_{\mu}A_{\nu}}\rangle_{\bf V}={1\over W({\bf V})} {\int W({\bf V},{\bf
A})A_{\mu}A_{\nu}d^{2}{\bf A}}
\label{A2V1}
\end{equation}
According to equation (\ref{WW1}), it can be written
\begin{equation}
W({\bf V})\langle {A_{\mu}A_{\nu}}\rangle_{\bf V}={1\over 16\pi^{4}} \int
A({\mb\rho},{\mb\sigma})   e^{-i({\mb\rho}{\bf V}+{\mb\sigma}{\bf A})}
A_{\mu}A_{\nu}d^{2}{\mb\rho}d^{2}{\mb\sigma}d^{2}{\bf A}
\label{A2V2}
\end{equation}
or, equivalently,
\begin{equation}
W({\bf V})\langle {A_{\mu}A_{\nu}}\rangle_{\bf V}=-{1\over 16\pi^{4}} \int
A({\mb\rho},{\mb\sigma}) {\partial^{2}\over \partial
\sigma^{\mu}\partial\sigma^{\nu}}\biggl \lbrace e^{-i({\mb\rho}{\bf
V}+{\mb\sigma}{\bf A})}\biggr \rbrace  d^{2}{\mb\rho}d^{2}{\mb\sigma}d^{2}{\bf
A}  
\label{A2V3}
\end{equation}
Integrating by parts, we obtain
\begin{equation}
W({\bf V})\langle {A_{\mu}A_{\nu}}\rangle_{\bf V}=-{1\over 16\pi^{4}} \int
{\partial^{2}A\over \partial
\sigma^{\mu}\partial\sigma^{\nu}}({\mb\rho},{\mb\sigma}) e^{-i({\mb\rho}{\bf
V}+{\mb\sigma}{\bf A})} d^{2}{\mb\rho}d^{2}{\mb\sigma}d^{2}{\bf A} 
\label{A2V4}
\end{equation}
Using the identity (\ref{delta}), we can readily carry out the integration on
${\bf A}$  and ${\mb\sigma}$ to finally obtain
\begin{equation}
W({\bf V})\langle {A_{\mu}A_{\nu}}\rangle_{\bf V}=-{1\over 4\pi^{2}} \int
{\partial^{2}A\over \partial \sigma^{\mu}\partial\sigma^{\nu}}({\mb\rho},{\bf
0}) e^{-i{\mb\rho}{\bf V}}  d^{2}{\mb\rho}
\label{A2V5}
\end{equation} 
Since the characteristic function $A({\mb\rho},{\mb\sigma})$ is
isotropic  in each of its variables [see equations (\ref{AA6})(\ref{CC2})], the
tensor $\langle {A_{\mu}A_{\nu}}\rangle_{\bf V}$  is diagonal and can be
expressed as
\begin{equation}
\langle {A_{\mu}A_{\nu}}\rangle_{\bf V}={1\over 2} \langle A^{2}\rangle_{\bf V} 
\delta_{\mu\nu}
\label{A2V6}
\end{equation}
where $\langle A^{2}\rangle_{\bf V}$ is given by
\begin{equation}
W({\bf V})\langle A^{2}\rangle_{\bf V}=-{1\over \pi^{2}} \int {\partial
A\over\partial (\sigma^{2})}({\mb\rho},{\bf 0})e^{-i{\mb\rho}{\bf V}}
d^{2}{\mb\rho}
\label{A2V7}
\end{equation}
We need therefore the behavior of $A({\mb\rho},{\mb\sigma})$ for  $|{\mb
\sigma}|\rightarrow 0$, or according to (\ref{AA6}) the behavior of
$C({\mb\rho},{\mb\sigma})$ for $|{\mb \sigma}|\rightarrow 0$.  Introducing the
function $C({\mb\rho})$ defined in section \ref{sec_WV} [see equation
(\ref{C5})], and using equation (\ref{CC2}),  we can write
\begin{equation}
C({\mb \rho},{\mb \sigma})=C({\mb \rho})+F({\mb \rho},{\mb \sigma})
\label{CC5}
\end{equation}
with
\begin{equation}
F({\mb \rho},{\mb \sigma})= {\gamma^{2}\over 2\pi}\int_{0}^{+\infty}
J_{0}({\rho}{\Phi}) (1- e^{-{1\over 4}\pi n\ln N\sigma^{2}\Phi^{4}} ){d\Phi\over
\Phi^{3}}
\label{FF1}
\end{equation}
We have let $R\rightarrow\infty$ since the  integral is  convergent when
$\Phi\rightarrow 0$. In terms of this new function, the  expression for $\langle
A^{2}\rangle_{\bf V}$ can be rewritten   
\begin{equation}
W({\bf V})\langle A^{2}\rangle_{\bf V}={n\over \pi^{2}} \int e^{-n
C({\mb\rho})}{\partial F\over\partial (\sigma^{2})}({\mb\rho},{\bf
0})e^{-i{\mb\rho}{\bf V}} d^{2}{\mb\rho} 
\label{A2V8}
\end{equation}
It is shown in Appendix \ref{sec_formula} that
\begin{equation}
{\partial F\over\partial (\sigma^{2})}({\mb\rho},{\bf 0})={\pi n\over
4}\gamma^{2}\ln N \delta({\mb \rho})
\label{Fdelta}
\end{equation}
where $\delta$ stands for Dirac $\delta$-function. Substituting the foregoing
expression in  equation (\ref{A2V8}), we obtain
\begin{equation}
W({\bf V})\langle A^{2}\rangle_{\bf V}={n^{2}\gamma^{2}\over 4\pi}\ln N
\label{A2V9}
\end{equation}
Therefore, $\langle A^{2}\rangle_{\bf V}$ behaves like the inverse of the
velocity  distribution $W({\bf V})$. Combining equations (\ref{A2V9})
(\ref{Wgauss}) and (\ref{Wt6}), we obtain
\begin{equation}
\langle A^{2}\rangle_{\bf V}={n^{3}\gamma^{4}\over 16\pi}(\ln N)^{2}e^{4\pi
V^{2}\over n \gamma^{2}\ln N}\quad (V\lesssim V_{crit}(N))
\label{A2V10}
\end{equation}
\begin{equation}
\langle A^{2}\rangle_{\bf V}=n\pi\ln N  V^{4}\quad (V\gtrsim V_{crit}(N))
\label{A2V11}
\end{equation}
For $V\rightarrow\infty$, $\langle A^{2}\rangle_{\bf V}$ behaves like $V^{4}$.
This  result finds a simple physical interpretation in the nearest neighbor
approximation (see section \ref{sec_NNA}). 

By a similar procedure, we can calculate the variance of the velocity for an
assigned rate of change. We find: 
\begin{equation}
\langle V^{2}\rangle_{\bf A}={n\gamma^{2}\over 4\pi}\ln N \quad (A\lesssim
A_{crit}(N))
\label{V2A5}
\end{equation}
\begin{equation}
\langle V^{2}\rangle_{\bf A}={2 A\over \pi\sqrt{n\ln N}}\quad (A\gtrsim
A_{crit}(N)) 
\label{V2A11}
\end{equation}
For moderate values of $A$, the variance $\langle V^{2}\rangle_{\bf A}$ is
independent from $A$ and  coincides with formula (\ref{variance}). For large
fluctuations, the result (\ref{V2A11}) can be recovered in the nearest neighbor
approximation (see section \ref{sec_NNA}).  The crossover between the two
distributions (\ref{V2A5}) (\ref{V2A11}) occurs at
\begin{equation}
A_{crit}(N)\sim {n^{3/2}\gamma^{2}\over 8}(\ln N)^{3/2} 
\label{Acrit}
\end{equation}
a formula which  also has a clear physical interpretation within the nearest
neighbor approximation.

\section{The speed of fluctuations and the diffusion coefficient}
\label{sec_speed}

\subsection{The mean lifetime of a velocity fluctuation $V$}
\label{sec_meanlifeV}

On the basis of very general considerations (see, e.g., \cite{weiss} Weiss {\it
et al.} 1998), we would expect that the typical duration of the velocity
fluctuations be
\begin{equation}
T_{typ}\sim {d\over\sqrt{\langle V^{2}\rangle}}
\label{Ttypgen}
\end{equation}
This corresponds to the time needed by a vortex with typical velocity
$\sqrt{\langle V^{2}\rangle}$  to cross  the interparticle distance $d\sim
n^{-1/2}$. Using expression (\ref{variance}) for $\langle V^{2}\rangle$, we find
\begin{equation}
T_{typ}\sim {1\over n\gamma\sqrt{\ln N}}
\label{Ttypexp}
\end{equation}
Apart from the logarithmic correction, this formula can be obtained by direct
dimensional analysis.  However, if we define the mean lifetime of a state $V$ by
the formula
\begin{equation}
T({V})= {|{\bf V}|\over\sqrt{\langle A^{2}\rangle_{\bf{V}}}},
\label{TVdef}
\end{equation}
the theory developed along the previous lines enables us to obtain a more
precise characterization  of the speed of fluctuations depending on their
intensity. This definition is consistent with  the definition introduced by
\cite{cn42,cn43} Chandrasekhar \& von Neumann (1942,1943) in the context of
stellar dynamics. Of course,  equation (\ref{TVdef}) is just an order of
magnitude but it should reasonably  well account for the dependence on the speed
of fluctuations with $V$.  Using (\ref{A2V10}) and (\ref{A2V11}) we find
explicitly
\begin{equation}
T({ V})= {4\sqrt{\pi}V\over n^{3/2}\gamma^{2}\ln N} e^{-{2\pi\over
n{\gamma^{2}}\ln N}V^{2}}\quad (V\lesssim V_{crit}(N))    
\label{TV1}
\end{equation}
\begin{equation}
T({V})= {1\over \sqrt{\pi n \ln N}}{1\over V}  \quad (V\gtrsim V_{crit}(N))    
\label{TV2}
\end{equation}
For weak and large fluctuations, $T(V)$ decreases to zero like $V$ and $V^{-1}$
respectively.  These asymptotic behaviors are consistent with the theory
developed by \cite{smoluchowski} Smoluchowski (1916) in his general
investigation on fluctuation phenomena (see section \ref{sec_smolu}). These
results (and those of section \ref{sec_statacceleration}) should be all the more
valid if ${\bf V}$ is calculated at a fixed point. By contrast, if ${\bf V}$
denotes the velocity experienced by a test vortex, we expect some discrepancies
at large $V$'s due to the formation of pairs. In such a case, the correlation
time can be extremely long (in particular for binaries).

The average duration of the fluctuations is defined by
\begin{equation}
\langle T\rangle =\int_{0}^{+\infty} T(V)W(V)2\pi V dV
\label{Taverage}
\end{equation}
To leading order in $\ln N$ (i.e. extending equations (\ref{Wgauss}) and
(\ref{TV1}) to all $V$'s), we obtain 
\begin{equation}
\langle T\rangle= {4\over 3}\biggl ({\pi\over 6}\biggr )^{1/2} {1\over
n\gamma\sqrt{\ln N}}
\label{Tcore}
\end{equation} 
in good agreement with the estimate (\ref{Ttypexp}) based on general physical
grounds.

\subsection{The diffusion coefficient}
\label{sec_diffusion}

According to the previous discussion, we can characterize the fluctuations of
the velocity of a point vortex (or a passive particle) by two functions: a
function $W({\bf V})$ which governs the occurrence of the velocity  ${\bf V}$
and a function $T(V)$ which determines the typical time during which the vortex
moves with this velocity. Since the velocity fluctuates on a typical time
$T_{typ}={d\over \sqrt{\langle V^{2}\rangle}}$ which is much smaller than the
dynamical time $T_{D}={R\over \sqrt{\langle V^{2}\rangle}  }$ needed by the
vortex to cross the entire domain, the motion of the vortex will be essentially
stochastic. If we denote by $P({\bf r},t)$ the probability density that the
particle be found in ${\bf r}$ at time $t$, then $P({\bf r},t)$ will satisfy the
diffusion equation 
\begin{equation}
{\partial P\over\partial t}=D\Delta P
\label{diffeq}
\end{equation}
If the particle is at ${\bf r}={\bf r}_{0}$ at time $t=0$, the solution of
equation (\ref{diffeq}) is clearly 
\begin{equation}
P({\bf r},t|{\bf r}_{0})={1\over 4\pi D t}e^{-{|{\bf r}-{\bf r}_{0}|^{2}\over 4
D t}}
\label{sol}
\end{equation}
where $D$ is the diffusion coefficient. The mean square displacement that  the
particle is expected to suffer during an interval of time $\Delta t$ large  with
respect to the fluctuation time $T_{typ}$, is
\begin{equation}
\langle (\Delta {\bf r})^{2}\rangle =4 D \Delta t
\label{rms}
\end{equation}
We can obtain another expression for $\langle (\Delta {\bf r})^{2}\rangle$ in
terms of the functions $W({\bf V})$ and $T(V)$ defined in the previous sections.
Indeed, dividing the interval 
\begin{equation}
\Delta {\bf r}=\int_{t}^{t+\Delta t}{\bf V}(t') dt'
\label{div}
\end{equation}
into a succession of discrete increments in position with amount $T(V_{i}){\bf
V}_{i}$,  we readily establish that
\begin{equation}
\langle (\Delta {\bf r})^{2}\rangle = \langle T(V)V^{2}\rangle\Delta t
\label{rmsbis}
\end{equation}
Combining equations (\ref{rms}) and (\ref{rmsbis}) we obtain an alternative
expression for the diffusion coefficient in the form
\begin{equation}
D={1\over 4}\int T(V)W({\bf V})V^{2} d^{2}{\bf V}
\label{diff2}
\end{equation}
Substituting for $T(V)$ and $W({\bf V})$ in the foregoing expression, we obtain
to leading order in $\ln N$:
\begin{equation}
D={1\over 72}\biggl ({6\over\pi}\biggr )^{1/2}\gamma\sqrt{\ln N} 
\label{Dexp}
\end{equation} 
We should not give too much credit to the numerical factor appearing in equation
(\ref{Dexp}) since the definition (\ref{TVdef}) of $T(V)$ is just an order of
magnitude. Note that the functional form of $D$ is consistent with the
expression
\begin{equation}
D\sim T_{typ}\langle V^{2}\rangle \sim\gamma\sqrt{\ln N} 
\label{Dtyp}
\end{equation}
that one would expect on general physical grounds. \cite{weiss} Weiss {\it
et al.} (1998) have proposed to describe the motion of the vortices by  a
Ornstein-Uhlenbeck (OU) process with ``friction'' $-{\bf V}/<T>$ to take into
account the finite decorrelation time of the system.  The function $T(V)$
introduced in the present article could be used instead of $\langle T\rangle$ to
take into account the dependence of the decorrelation time with the strength $V$
of the fluctuations.

The present theory ignores the formation of pairs since we have formally
introduced a cut-off at large $V$'s. This cut-off is justified for binaries
since, as we have already explained, the relevant velocity to consider is the
velocity of the center of vorticity not the velocity of the individual vortices
engaged in the pair. This is not the case for dipoles which can make relatively
long jumps from one point to another with an almost ballistic
motion.\cite{weiss} Weiss {\it et al.} (1998) have proposed to interpret these
jumps in terms of L\'evy walks responsible for anomalous diffusion. Therefore,
the present theory should  provide reliable results only for relatively short
times. For times $t\gg T_{D}$, the pairs must be taken into account and
anomalous diffusion will ensue. The case of passive particles is not so
different. A passive particle is advected by the other vortices but has no
influence on their motion. However, passives  can get trapped in the vicinity of
a vortex undergoing fast dipolar motion and  experience L\'evy walks also. 

In \cite{sire}, we have actually made the connection between  L\'evy walks and
anomalous diffusion more explicit, by showing that the flight time p.d.f. has
a slow power-law decaying tail with an exponent that can be related to the
number of vortices decay exponent $\xi$ (see next section).

\subsection{Application to 2D decaying turbulence}
\label{sec_anomalous} 

Let us consider a collection of vortices of size $a$, vorticity $\omega$ and
density $n$. Due to merging events, their size increases with time as the
density decreases. The typical core vorticity $\omega$ remains constant during
the course of the evolution as suggested in \cite{carnevale} Carnevale {\it et
al.} (1991) \cite{benzi} Benzi {\it et al.} (1992) and \cite{weissmcwilliams}
Weiss {\it et al.} (1993), and observed experimentally by \cite{hansen} Hansen
{\it et al.} (1998).  These authors found that the density decreases as $n\sim
t^{-\xi}$  with $\xi\approx 0.7$. As the energy $E\sim N\omega^{2}a^{4}$ is
conserved throughout the merging process, the typical vortex radius $a\sim
t^{\xi/4}$.  Since the average distance between vortices, of order $d\sim
t^{\xi/2}$ increases more rapidly than their size, the point vortex model should
provide increasing accuracy. We can therefore expect that the vortices will
diffuse with a coefficient $D\sim \gamma$ [see equation (\ref{Dexp})] where
$\gamma\sim \omega a^{2}$ is their circulation (we ignore logarithmic
corrections). If the diffusion coefficient were constant, then the dispersion of
the vortices  
\begin{equation}
\langle r^{2} \rangle\sim D t
\label{disp}
\end{equation}
would increase linearly with time as in ordinary  Brownian motion. However,
since $D$ varies with time according to
\begin{equation}
D\sim \omega a^{2} \sim t^{\xi/2}
\label{Dtime}
\end{equation}
we expect anomalous diffusion, i.e. 
\begin{equation}
\langle r^{2} \rangle\sim t^{\nu}
\label{anomal}
\end{equation}
with $\nu\neq 1$. Substituting equation (\ref{Dtime}) into equation (\ref{disp})
we obtain the following relation between $\nu$  and $\xi$:
\begin{equation}
\nu =1+{\xi\over 2}
\label{nuxi}
\end{equation}     

This expression differs from formula $(19)$ of \cite{hansen} Hansen {\it et al.}
(1998) because their estimate of $D$ is different. These authors estimate the
diffusion coefficient by $D\sim \tau_{merg}\langle V^{2}\rangle$ where
$\tau_{merg}$ is the average time between two successive mergings of a given
vortex. By a simple cross section argument, they obtain $\tau_{merg}\sim {1\over
n\sqrt{\langle V^{2}\rangle} a}$ which is larger than $\tau_{fluct}\sim {d\over
\sqrt{\langle V^{2}\rangle}}$ by a factor ${d\over a}$. This shows that the
merging time is not the relevant correlation time to consider in the diffusion
process. In fact, using $\xi\approx 0.7$, formula (\ref{nuxi}) leads to
$\nu\approx 1.35$ in better agreement  with the experimental value  $\nu\approx
1.3$ ( $\nu\approx 1.4$ for passive particles) than their relation $\nu
=1+{3\xi\over 4}$ which leads to $\nu\approx 1.53$. Formula (\ref{nuxi}) is also
in perfect agreement with the numerical simulations of \cite{sire} Sire \&
Chavanis (1999).

\section{The nearest neighbor approximation}
\label{sec_NNA}

\subsection{The importance of the nearest neighbor}
\label{sec_importance}

The velocity ${\bf V}$ and the acceleration ${\bf A}$ experienced by a test
vortex (or occurring at a fixed point) are the sum of $N$ random variables
${\mb\Phi}_{i}$ and ${\mb\psi}_{i}$ produced by all the vortices present in the
system [see equations (\ref{vtot}) and (\ref{A})]. In each sum, the highest term
is due to the nearest neighbor, at an average distance $d\sim n^{-1/2}$ from
the point under consideration. This single vortex creates a typical velocity and
acceleration
\begin{equation}
V^{2}_{n.n}\sim  \biggl ({\gamma\over 2\pi d}\biggr )^{2} \sim {\gamma^{2}\over
4 \pi^{2}} {N\over \pi R^{2}}
\label{V2nn}
\end{equation}
\begin{equation}
A^{2}_{n.n}\sim {\overline{v^{2}}} \biggl ({\gamma\over 2\pi d^{2}}\biggr )^{2}
\sim {\gamma^{2}\over 4 \pi^{2}}{\overline{v^{2}}} \biggl ({N\over \pi
R^{2}}\biggr )^{2}
\label{A2nn}
\end{equation}
It is interesting to compare the contribution of the nearest neighbor with the
contribution of the other $N-1$ vortices. For that purpose, we estimate the
typical value of $V$ and $A$ produced by {\it all} the vortices by
\begin{equation}
V^{2}\sim N\biggl\langle {\gamma^{2}\over 4 \pi^{2} r^{2}}\biggr\rangle \sim
N\int_{|{\bf r}|=d}^{R} \tau({\bf r}) {\gamma^{2}\over 4\pi^{2}r^{2}}d^{2}{\bf
r}\sim {\gamma^{2}\over 4\pi}{N\over \pi R^{2}}\ln N 
\label{V2typ}
\end{equation}
\begin{equation}
A^{2}\sim N \overline{v^{2}}  \biggl\langle {\gamma^{2}\over 4 \pi^{2}
r^{4}}\biggr\rangle \sim N  \overline{v^{2}}   \int_{|{\bf r}|=d}^{R} \tau({\bf
r}) {\gamma^{2}\over 4\pi^{2}r^{4}}d^{2}{\bf r}\sim {\gamma^{2}\over
4\pi}\overline{v^{2}} \biggl ({N\over \pi R^{2}}\biggr )^{2} 
\label{A2typ}
\end{equation}

If the variance of ${\mb \Phi}$ and ${\mb\psi}$ were finite, the Central Limit
Theorem would be applicable and the variables $V$ and $A$ would scale like
$\sqrt{N}$. In that case, none of the terms in the sums (\ref{vtot}) (\ref{A})
would have a dominant contribution and the scaling $\sqrt{N}$ would simply
reflect the {\it collective} behavior of the system.  This is not the case
however in the present situation since  the variance of ${\mb\Phi}$ and
${\mb\psi}$ diverge. The variance of ${\mb \psi}$ diverges algebraically and,
thus, the acceleration produced by all the vortices is dominated by the
contribution of the nearest neighbor. This {\it individual} nature is a specific
and striking property of a L\'evy law. For a L\'evy law, the sum of $N$ random
variables behaves like the largest term (compare equations (\ref{A2typ}) and
(\ref{A2nn})).  The case of the velocity is particular because the variance of
${\mb\Phi}$ diverges only logarithmically. First considering  equation
(\ref{V2typ}), and neglecting the logarithmic correction, we observe that the
velocity produced by all the vortices behaves like $\sqrt{N}$, as though the
Central Limit Theorem were applicable. However, comparing with equation
(\ref{V2nn}), we note that $\sqrt{N}$ is also the scaling of the largest term in
the sum. Therefore, the velocity has a behavior which is intermediate between
Gaussian and L\'evy laws, as remarked earlier. In a sense, we can consider that
the velocity is dominated by the contribution of the nearest neighbor and that
collective effects are responsible for logarithmic corrections. 

\subsection{The distribution due to the nearest neighbor}
\label{sec_distribution}

In light of the previous discussion, it is interesting to analyze in more detail
the distribution of the velocity and acceleration produced by the nearest
neighbor. For that purpose, we  must first determine the probability
$\tau_{n.n}(r)dr$ that the position of the nearest neighbor occurs between $r$
and $r+dr$. Clearly,  $\tau_{n.n}(r)dr$ is equal to the probability that no
vortices exist interior to $r$ times the probability that a vortex (any) exists
in the annulus between $r$ and $r+dr$. Therefore, it must satisfy an equation of
the form:
\begin{equation}
\tau_{n.n}(r)dr=\biggl (1-\int_{0}^{r}\tau_{n.n}(r')dr'\biggr ) n 2\pi r dr
\label{tnnrel}
\end{equation}
where $n={N\over \pi R^{2}}$ denote the mean density of vortices in the disk.
Differentiating with respect to $r$ we obtain 
\begin{equation}
{d\over dr}\biggl\lbrack{\tau_{n.n}(r)\over 2\pi n r}\biggl\rbrack
=-\tau_{n.n}(r) 
\label{tnndiff}
\end{equation} 
This equation is readily integrated with the condition $\tau_{n.n}(r)\sim 2\pi n
r$ as $r\rightarrow 0$, and we find
\begin{equation}
\tau_{n.n}(r)={2\pi n} r e^{-{\pi n}r^{2}}
\label{tnn}
\end{equation}
This is the distribution of the nearest neighbor in a random distribution of
particles. From this formula, we can obtain the exact value for the ``average
distance'' $d$ between vortices. By definition,  
\begin{equation}
d=\int_{0}^{+\infty} \tau_{n.n}({ r})r d{ r}
\label{ddef}
\end{equation}
Hence
\begin{equation}
d={1\over 2\sqrt{n}}
\label{d}
\end{equation}

The probability of finding the nearest neighbor in ${\bf r}$ with velocity
${\bf v}$ is 
\begin{equation}
\tau_{n.n}({\bf r},{\bf v})= n e^{-{\pi n}r^{2}} \times {4\over n\gamma^{2}\ln
N}e^{-{4\pi\over n\gamma^{2}\ln N}v^{2}}
\label{tnnrv}
\end{equation}
If we assume that the velocity ${\bf V}$ and acceleration ${\bf A}$ are entirely
due to the nearest neighbor, then 
\begin{equation}
W_{n.n}({\bf V},{\bf A})d^{2}{\bf V}d^{2}{\bf A}=\tau_{n.n}({\bf r},{\bf
v})d^{2}{\bf r}d^{2}{\bf v}
\label{WnnVA}
\end{equation}  
with
\begin{equation}
{\bf V}=-{\gamma\over 2\pi}{{\bf r}_{\perp}\over r^{2}}
\label{Vsol}
\end{equation}
\begin{equation}
{\bf A}=-{\gamma\over 2\pi}\biggl ( {{\bf v}_{\perp }\over r_{}^{2}}-
{2({\bf r}\cdot {\bf v}){\bf r}_{\perp }\over r^{4}}\biggr ) 
\label{Asol}
\end{equation}
Since the Jacobian of the transformation $\lbrace {\bf r},{\bf
v}\rbrace\rightarrow \lbrace {\bf V},{\bf A}\rbrace$ is 
\begin{equation}
\biggl |\biggl | {\partial ({\bf r},{\bf v})\over\partial ({\bf V},{\bf
A})}\biggr |\biggr |={\gamma^{4}\over 16\pi^{4} V^{8}}
\label{jacnn}
\end{equation}
we obtain:
\begin{equation}
W_{n.n}({\bf V},{\bf A})={\gamma^{2}\over 4\pi^{4}\ln N}{1\over
V^{8}}e^{-{n\gamma^{2}\over 4\pi V^{2}}}e^{-{A^{2}\over n\pi\ln N V^{4}}}
\label{WnnVAexp}
\end{equation}
where we have used 
\begin{equation}
r={\gamma\over 2\pi V}\qquad {\rm and}\qquad v={\gamma\over 2\pi}{A\over V^{2}}
\label{rVnew}
\end{equation}
The nearest neighbor approximation is expected to give relevant results only
for large values of the velocity and the acceleration. Thus, we can make the
additional approximation 
\begin{equation}
W_{n.n}({\bf V},{\bf A})={\gamma^{2}\over 4\pi^{4}\ln N}{1\over
V^{8}}e^{-{A^{2}\over n\pi\ln N V^{4}}}
\label{WnnVAapp}
\end{equation}

Integrating on the acceleration, we find
\begin{equation}
W_{n.n}({\bf V})={n\gamma^{2}\over 4\pi^{2} V^{4}}
\label{WnnV}
\end{equation}
in perfect agreement with equation (\ref{Wt6}) valid for $V\gtrsim V_{crit}$.
This shows that the algebraic tail of the velocity distribution is produced by
the nearest neighbor. This is characteristic of a L\'evy law. On the other hand,
for $V<V_{crit}$, the velocity distribution is Gaussian as if the Central Limit
Theorem were applicable. Once again, the simultaneous occurrence of collective
and individual behaviors is a manifestation of the very peculiar nature of an
interaction in $r^{-1}$ in two dimensions. 

Integrating on the velocity, we find
\begin{equation}
W_{n.n}({\bf A})={\gamma^{2}n^{3/2}\sqrt{\ln N}\over 16\pi A^{3}}
\label{WnnA}
\end{equation}
in perfect agreement with the asymptotic behavior of the Cauchy distribution
(\ref{WA4}). We also establish that 
\begin{equation}
\langle A^{2}\rangle_{\bf V}=n\pi \ln N V^{4}
\label{A2Vnn}
\end{equation}
\begin{equation}
\langle V^{2}\rangle_{\bf A}={2 A\over \pi\sqrt{n\ln N}}
\label{V2Ann}
\end{equation}
in complete agreement with formulae (\ref{A2V11}) and (\ref{V2A11}).

\subsection{The application of Smoluchowski (1916) theory}
\label{sec_smolu}

In the nearest neighbor approximation, the duration of the velocity fluctuations
can be deduced from the theory of \cite{smoluchowski} Smoluchowski (1916)
concerning the persistence of fluctuations. This approach was used by
\cite{chandr} Chandrasekhar (1941) in his elementary analysis of the
fluctuations of the gravitational field. An account of Smoluchowski theory can
be found in \cite{kandrup} Kandrup (1980). In the case of point vortices, it
leads to the formula 
\begin{equation}
T(V)={\gamma\over 4\overline{v}}{V\over 
{n\gamma^{2}\over 4\pi}+V^{2}}
\label{TVnn}
\end{equation} 
where $V={\gamma\over 2\pi r}$ is the velocity due to the most proximate vortex,
at a distance $r$ from the point under consideration. The Smoluchowski formula
(\ref{TVnn}) has the same asymptotic behaviors as (\ref{TV1}) and (\ref{TV2}).
These asymptotic behaviors have a clear physical meaning.  When $r={\gamma\over
2\pi V}$ is small, corresponding to large velocities, it is highly improbable
that another vortex will enter the disk of radius $r$ before long. By contrast,
on a short time scale $T\sim {r\over\overline{v}}\sim
{\gamma\over\overline{v}V}$, the vortex will have left the disk. When
$r={\gamma\over 2\pi V}$ is large, corresponding to small velocities,  the
probability that the vortex will remain alone in the disk is low. The
characteristic time before another vortex enters the disk varies like the
inverse of the number of vortices expected to be present in the disk, i.e.
$T\sim {r\over\overline{v}}{1\over n\pi r^{2}}\sim {V\over n \gamma
\overline{v}}$. The demarcation between weak and strong fluctuations corresponds
to $V\sim \gamma n^{1/2}$, i.e. to the velocity produced by a vortex distant
$n^{-1/2}$ from the point under consideration.

\section{Conclusion}

In this paper, we have analyzed in detail the statistics of velocity
fluctuations produced by a  random distribution of point vortices. We have
determined the velocity distribution and the speed of fluctuations. We have also
shown how some of the results can be understood in the nearest neighbor
approximation. In a future study (in preparation), we shall be concerned with
the spatial and temporal correlations of the velocity. Our results should be
accurate if the velocity is calculated at a fixed point. However, there should
be substantial discrepancies at large velocities if ${\bf V}$ now represents the
velocity experienced by a point vortex. This is due to the formation of pairs
(binaries or dipoles) when two vortices come into contact. If we ignore these
pairs, the motion of the vortices is purely diffusive and we determined the
functional form of the diffusion coefficient (up to a numerical factor). In the
case of real vortices, with a finite core, the formation of binaries is replaced
by merging events and the number of vortices decreases with time. This results
in anomalous diffusion. We proposed a relationship between the exponent of
anomalous diffusion $\nu$ and the exponent $\xi$ which characterizes the decay
of the vortex density.

Our study was directly inspired by the work of \cite{cn42,cn43} Chandrasekhar \&
von Neumann (1942,1943) concerning the statistics of the gravitational field
arising from a random distribution of stars. Of course, the results differ due
to the different nature of the interactions but the mathematical structure of
the problem is the same. This is another example of the analogy between
two-dimensional vortices and stellar systems investigated by
\cite{chav96,chav98b} Chavanis (1996,1998b) and \cite{csr96} Chavanis {\it et
al.} (1996). Likewise in the stellar context, the formation of binaries alters
the results of the stochastic analysis  at large field strengths.   

\acknowledgements 

We are grateful to Jane Basson for useful comments on the manuscript. We thank
the referee for providing us with the reference of Jim\'enez paper. After the
submission of this article we have been aware of the work by Kuvshinov \& Schep
(to appear in Phys. Rev. Lett.) on the statistics of point vortex systems.

\newpage
\appendix

\section{Derivation of formula (91).}
\label{sec_formula}

By definition,
\begin{equation}
F({\mb \rho},{\mb \sigma})= {\gamma^{2}\over 2\pi}\int_{0}^{+\infty}
J_{0}({\rho}{\Phi}) (1- e^{-{1\over 4}\pi n\ln N\sigma^{2}\Phi^{4}}
){d\Phi\over \Phi^{3}}
\label{AF}
\end{equation}
For ${\mb \rho}={\bf 0}$, we have already found [see equations (\ref{CC3}) and
(\ref{CC4})] that
\begin{equation}
F({\bf 0},{\mb \sigma})= {\gamma^{2}\over 8}\sqrt{n\ln N}\sigma
\label{AF0}
\end{equation}
Therefore:
\begin{equation}
{\partial F\over\partial (\sigma^{2})}({\mb\rho},{\bf 0})= +\infty 
\quad {\rm if} \quad {\mb \rho}={\bf 0}
\label{ADF0}
\end{equation}
For ${\mb \rho}\neq {\bf 0}$, we can make the change of variable $z=\rho\Phi$
in equation (\ref{AF}). This yields
\begin{equation}
F({\mb \rho},{\mb \sigma})= {\gamma^{2}\rho^{2}\over 2\pi}\int_{0}^{+\infty}
J_{0}(z) (1- e^{-{1\over 4}\pi n\ln N{\sigma^{2}\over\rho^{4}}z^{4}} ){dz\over
z^{3}}
\label{AFbis}
\end{equation}
We have therefore to determine the behavior of the function 
\begin{equation}
f(p)= \int_{0}^{+\infty} J_{0}(z)(1- e^{-p z^{4}} ){dz\over z^{3}}
\label{Af}
\end{equation}
as $p\rightarrow 0$. Clearly, it is not possible to expand the quantity $$1-
e^{-p z^{4}}$$ which occurs under the integral sign as a power series of $p
z^{4}$ and evaluate the integral term by term. However, writing the Bessel
function in the form
\begin{equation}
J_{0}(z)={1\over \pi}{\cal R}_{e}\int_{-1}^{+1} e^{i z t }{dt\over
\sqrt{1-t^{2}}}
\label{Abesint}
\end{equation} 
and regarding $z$ and $t$ as complex variables, it is possible to choose
integration paths along which this expansion will converge. Using the contours
introduced in  section \ref{sec_tail}, the function $f(p)$ can be rewritten  
\begin{equation}
f(p)={1\over \pi}{\cal R}_{e}\int_{\tau} {dt\over \sqrt{1-t^{2}}}
\int_{\zeta_{\omega_{t}}}e^{i z t } \biggl (1- e^{-p z^{4}}\biggr ){dz\over
z^{3}}
\label{Afp}
\end{equation}
We readily verify that the real parts of $i z t $ and $-p z^{4}$ are always
negative, so the convergence of equation (\ref{Af}) is not disturbed. With these
new contours, it is now possible to expand the integrand in a power series of
$pz^{4}$ and integrate term by term. For our purposes, it is only necessary to
consider the term of first order in this expansion
\begin{equation}
f(p)=p {1\over \pi}{\cal R}_{e}\int_{\tau} {dt\over \sqrt{1-t^{2}}} 
\int_{\zeta_{\omega_{t}}}e^{i z t } z dz +O(p^{2})
\label{Afp1}
\end{equation} 
Therefore
\begin{equation}
f'(0)= {1\over \pi}{\cal R}_{e}\int_{\tau} {dt\over \sqrt{1-t^{2}}}
\int_{\zeta_{\omega_{t}}}e^{i z t } z dz 
\label{Afprime0}
\end{equation}
The integration on $z$ can be carried out equivalently along the line
$\zeta_{\psi_{t}}$, defined in section \ref{sec_tail}, on which $i z t =-y$,
$y\ge 0$ real. We obtain
\begin{equation}
f'(0)=- {1\over \pi}{\cal R}_{e}\int_{-1}^{+1} {dt\over \sqrt{1-t^{2}}}
\int_{0}^{+\infty} e^{-y } {y\over t^{2}} dy 
\label{Afprime0bis}
\end{equation} 
where $t$ is a complex variable running along the semi-circle of radius unity
lying in the domain ${\cal I}_{m}(t)\ge 0$. Since
\begin{equation}
\int_{-1}^{+1} {dt\over t^{2}\sqrt{1-t^{2}}}=0 
\label{Aid}
\end{equation} 
we find $f'(0)=0$. Therefore
\begin{equation}
{\partial F\over\partial (\sigma^{2})}({\mb\rho},{\bf 0})= 0 
\quad {\rm if} \quad {\mb \rho}\neq {\bf 0}
\label{ADF0bis}
\end{equation}

To prove formula (\ref{Fdelta}), it remains to show that 
\begin{equation}
\int {\partial F\over\partial (\sigma^{2})}({\mb\rho},{\bf 0})d^{2}{\mb \rho}
={\pi n\over 4}\gamma^{2} \ln N
\label{Aint}
\end{equation}
For that purpose, we introduce the function
\begin{equation}
I(\sigma^{2})=\int { F}({\mb\rho},{\mb\sigma})d^{2}{\mb \rho}
\label{AI1}
\end{equation}
Substituting explicitly for $F({\mb\rho},{\mb\sigma})$ and introducing polar
coordinates, we get 
\begin{equation}
I(\sigma^{2})=\gamma^{2}\int_{0}^{+\infty}\rho
d\rho\int_{0}^{+\infty}J_{0}(\rho\Phi) (1- e^{-{1\over 4}\pi n\ln
N{\sigma^{2}}\Phi^{4}} ){d\Phi\over \Phi^{3}}
\label{AI2}
\end{equation}
Under this form, it is not possible to interchange the order of integration. An
alternative expression for $I(\sigma^{2})$ can be obtained along the following
lines. Writing 
\begin{equation}
I(\sigma^{2})=\gamma^{2}\int_{0}^{+\infty} d\rho \int_{0}^{+\infty}\rho\Phi
J_{0}(\rho\Phi)g(\Phi){d\Phi}
\label{AI3}
\end{equation}
where
\begin{equation}
g(\Phi)=(1- e^{-{1\over 4}\pi n\ln N{\sigma^{2}}\Phi^{4}}){1\over \Phi^{4}}
\label{Ag}
\end{equation}
and integrating by parts with the identity
\begin{equation}
xJ_{0}(x)={d\over dx}(xJ_{1}(x))
\label{Aidbis}
\end{equation}
we obtain
\begin{equation}
I(\sigma^{2})=-\gamma^{2}\int_{0}^{+\infty} d\rho \int_{0}^{+\infty}
J_{1}(\rho\Phi)\Phi g'(\Phi){d\Phi}
\label{AI4}
\end{equation}
It is now possible to interchange the order of integration. Since
\begin{equation}
\int_{0}^{+\infty} J_{1}(x) dx=1
\label{Aun}
\end{equation}
equation (\ref{AI4}) reduces to 
\begin{equation}
I(\sigma^{2})=-\gamma^{2}\int_{0}^{+\infty} g'(\Phi){d\Phi}=\gamma^{2}\ g(0)
\label{AI5}
\end{equation}
Hence
\begin{equation}
I(\sigma^{2})={\pi n\over 4}\gamma^{2}\ln N \sigma^{2}
\label{AI6}
\end{equation}
This formula is valid for any value of $\sigma$ but, for our purposes, we only
need the result
\begin{equation}
I'(0)={\pi n\over 4}\gamma^{2} \ln N
\label{AIprime0}
\end{equation}
Since, by definition,
\begin{equation}
I'(0)=\int {\partial F\over\partial (\sigma^{2})}({\mb\rho},{\bf 0})d^{2}{\mb
\rho}
\label{autre}
\end{equation}
we have proved equation (\ref{Aint}).

\section{Generalization to include a spectrum of circulations and an arbitrary
isotropic distribution of velocities}
\label{sec_generalization}

So far, we have assumed that the system was a ``vortex plasma''  consisting of
an equal number of vortices with circulation $+\gamma$ and $-\gamma$. We shall
now indicate how the previous results can be extended to include a spectrum of
circulations among the vortices. We shall also relax the assumption (\ref{df})
concerning the velocity distribution of the vortices and  generalize the results
of section \ref{sec_statacceleration} to any isotropic distribution  $\tau({\bf
v})=\tau(|{\bf v}|)$ of the velocities. Such a distribution can be written
conveniently in the form  
\begin{equation}
\tau({\bf v})=\int {\tau({\bf v}_{0})\over 2\pi v_{0}}\delta (v-v_{0})d^{2}{\bf
v}_{0}
\label{Btauv}
\end{equation}

If $\tau(\gamma)$ governs the distribution over the circulations, it is clear 
that equation (\ref{AA1}) has to be modified according to
\begin{equation}
A_{N}({\mb \rho},{\mb \sigma})=\biggl
(\int_{\gamma=-\infty}^{+\infty}\int_{|{\bf r}|=0}^{R}\int_{|{\bf
v}|=0}^{+\infty}\tau(\gamma) \tau({\bf r})\tau({\bf v})e^{i({\mb \rho}{\bf
\Phi}+{\mb \sigma}{\mb\psi})}d\gamma d^{2}{\bf r}d^{2}{\bf v}\biggr )^{N}
\label{BA1}
\end{equation}    
There is no a priori restriction on the function $\tau(\gamma)$ but we shall be
particularly interested in the case where the system is ``neutral'', i.e.
$\overline{\gamma}=\int\tau(\gamma)\gamma d\gamma=0$. It is only in this
circumstance that the velocity distribution (\ref{Btauv}) may be used.
Otherwise, there is a solid rotation of the system which adds to the dispersion
of the particles (see Appendix \ref{sec_nonneutral}). 

The expression (\ref{AA3}) for $A_{N}({\mb \rho},{\mb
\sigma})$ is now replaced by
\begin{eqnarray}
A_{N}({\mb \rho},{\mb \sigma})=\biggl ({1\over \pi R^{2}}
\int_{\gamma=-\infty}^{+\infty}\int_{|{\bf v}_{0}|=0}^{+\infty} \int_{|{\mb
\Phi}|={|\gamma|\over 2\pi R}}^{+\infty}\int_{|{\mb
\psi}|=0}^{+\infty}\nonumber\\ \times \tau(\gamma)  {\tau({\bf v}_{0})\over
2\pi v_{0}}\delta\biggl ({|\gamma|\psi\over 2\pi \Phi^{2}}-v_{0}\biggr )
e^{i({\mb \rho}{\bf \Phi}+{\mb \sigma}{\mb\psi})}{\gamma^{4}\over
16\pi^{4}\Phi^{8}} d\gamma d^{2}{\bf v}_{0}  d^{2}{\bf \Phi}d^{2}{\mb  \psi}
\biggr )^{N}
\label{BA2}
\end{eqnarray} 
Introducing polar coordinates, using identity (\ref{identity1}) and substituting
for
\begin{equation}
\delta\biggl ({|\gamma|\psi\over 2\pi \Phi^{2}}-v_{0}\biggr )={2\pi\Phi^{2}\over
|\gamma|}\delta \biggl (\psi-v_{0}{2\pi\Phi^{2}\over |\gamma|}\biggr )
\label{Bdelta}
\end{equation}
in equation (\ref{BA2}), we can easily integrate on $\psi$ to obtain
\begin{equation}
A_{N}({\mb \rho},{\mb \sigma})=\biggl ({1\over \pi
R^{2}}\int_{\gamma=-\infty}^{+\infty}\int_{|{\bf v}_{0}|=0}^{+\infty}\int_{|{\mb
\Phi}|={|\gamma|\over 2\pi R}}^{+\infty} \tau(\gamma) \tau({\bf v}_{0})
e^{i{\mb \rho}{\bf \Phi}} J_{0}\biggl ({2\pi\sigma\over
|\gamma|}v_{0}\Phi^{2}\biggr ) {\gamma^{2}\over 4\pi^{2}\Phi^{4}}d\gamma
d^{2}{\bf v}_{0} d^{2}{\mb\Phi}  \biggr )^{N}
\label{BA3}
\end{equation}  
It is readily verified that
\begin{equation}
{1\over \pi R^{2}}\int_{\gamma=-\infty}^{+\infty} \int_{|{\bf
v}_{0}|=0}^{+\infty} \int_{|{\mb \Phi}|={|\gamma|\over 2\pi R}}^{+\infty}
\tau(\gamma) \tau({\bf v}_{0})  {\gamma^{2}\over 4\pi^{2}\Phi^{4}}d\gamma
d^{2}{\bf v}_{0} d^{2}{\bf\Phi}   =1
\label{Bnorm}
\end{equation}   
Therefore, the expression for $A_{N}({\mb \rho},{\mb \sigma})$ can be rewritten
equivalently
\begin{eqnarray}
A_{N}({\mb \rho},{\mb \sigma})=\biggl (1-{1\over \pi R^{2}}
\int_{\gamma=-\infty}^{+\infty}    \int_{|{\bf v}|=0}^{+\infty}\int_{|{\mb
\Phi}|={|\gamma|\over 2\pi R}}^{+\infty}\nonumber\\ \times \tau(\gamma)
\tau({\bf v})       \biggl (1- e^{i{\mb \rho}{\bf \Phi}} J_{0}\biggl
({2\pi\sigma\over\gamma}v\Phi^{2}    \biggr )\biggr ) {\gamma^{2}\over
4\pi^{2}\Phi^{4}}d^{2}{\bf\Phi} d^{2}{\bf v} d\gamma \biggr )^{N}
\label{BA}
\end{eqnarray}
where we have written ${\bf v}$ instead of ${\bf v}_{0}$ as it is  a dummy
variable of integration. In the limit $N,R\rightarrow\infty$ with $n={N\over \pi
R^{2}}$ fixed, we obtain
\begin{equation}
A({\mb \rho},{\mb \sigma})=e^{-n C({\mb \rho},{\mb \sigma})}
\label{BA5}
\end{equation}   
with
\begin{equation}
C({\mb \rho},{\mb \sigma})= \int_{\gamma=-\infty}^{+\infty} \int_{|{\bf
v}|=0}^{+\infty} \int_{{|\gamma|\over 2\pi R}}^{+\infty}\tau(\gamma) \tau({\bf
v}_{0})    \biggl (1- J_{0}(\rho\Phi)J_{0}\biggl ({2\pi\sigma\over
|\gamma|}v\Phi^{2}    \biggr )   \biggr ) {\gamma^{2}\over 2\pi \Phi^{3}}
d\gamma d^{2}{\bf v}d{\Phi}  
\label{BC1}
\end{equation}

For ${\mb\sigma}={\bf 0}$, the function $C({\mb \rho},{\mb \sigma})$ reduces to:
\begin{equation}
C({\mb \rho})\simeq  {\overline{\gamma^{2}}\rho^{2}\over 16\pi}\ln N \quad
(\rho>0)
\label{BC3}
\end{equation} 
\begin{equation}
C({\mb \rho})\sim -{\overline{\gamma^{2}}\rho^{2}\over 8\pi}\ln \rho \quad
(\rho\rightarrow 0)
\label{BC2}
\end{equation} 
Therefore, the velocity distribution becomes
\begin{equation}
W({\bf V})={4\over n\overline{\gamma^{2}}\ln N}e^{-{4\pi\over
n\overline{\gamma^{2}}\ln N}V^{2}}\quad (V\lesssim V_{crit}(N))
\label{BWV1}
\end{equation} 
\begin{equation}
W({\bf V})={n\overline{\gamma^{2}}\over 4\pi^{2}V^{4}}   \quad (V\gtrsim
V_{crit}(N))
\label{BWV2}
\end{equation} 
with
\begin{equation}
V_{crit}(N)\sim \biggl ({n\overline{\gamma^{2}}\over 4\pi}\ln N\biggr )^{1/2} 
\label{BVcrit}
\end{equation} 
These results differ from (\ref{Wgauss}) (\ref{Wt6}) (\ref{Vcrit}) simply by the
substitution $\gamma^{2}\rightarrow \overline{\gamma^{2}}$ where
$\overline{\gamma^{2}}=\int \tau(\gamma)\gamma^{2}d\gamma$ is the average
enstrophy. 

Writing 
\begin{equation}
C({\mb \rho},{\mb \sigma})=C({\mb \rho})+F({\mb \rho},{\mb \sigma})
\label{BC4}
\end{equation}  
we get
\begin{equation}
F({\mb \rho},{\mb \sigma})= \int_{\gamma=-\infty}^{+\infty} \int_{|{\bf
v}|=0}^{+\infty}  \int_{0}^{+\infty} \tau(\gamma)  \tau({\bf v})
J_{0}(\rho\Phi)\biggl (1- J_{0}\biggl ({2\pi\sigma\over|\gamma|}v\Phi^{2}
\biggr )     \biggr ){\gamma^{2}\over 2\pi \Phi^{3}} d\gamma d^{2}{\bf v}
d{\Phi} 
\label{BF1}
\end{equation}  
Following the same steps as in Appendix \ref{sec_formula}, we find that
\begin{equation}
{\partial F\over\partial (\sigma^{2})}({\mb \rho},{\bf 0})=
\pi^{2}\overline{v^{2}}\delta ({\mb\rho})
\label{BDF}
\end{equation}
where 
\begin{equation}
\overline{v^{2}}=\int_{0}^{+\infty} \tau(|{\bf v}|)v^{2} 2\pi v dv
\label{Bv2}
\end{equation}
is the mean square velocity of the vortices for the isotropic distribution
$\tau(|{\bf v}|)$. Therefore, equations (\ref{A2V10}) and (\ref{A2V11}) are
modified according to
\begin{equation}
\langle A^{2}\rangle_{\bf V}={n^{2}\overline{\gamma^{2}}\over
4}\overline{v^{2}}\ln N e^{4\pi V^{2}\over n \overline{\gamma^{2}}\ln N}\quad
(V\lesssim V_{crit}(N))
\label{BA2V1}
\end{equation}
\begin{equation}
\langle A^{2}\rangle_{\bf
V}={4\pi^{2}\overline{v^{2}}\over\overline{\gamma^{2}}}V^{4}\quad
(V\gtrsim V_{crit}(N))
\label{BA2V2}
\end{equation}
and equations (\ref{TV1}) (\ref{TV2}) according to
\begin{equation}
T(V) ={2V\over n\sqrt{\overline{\gamma^{2}}}\sqrt{\ln
N}\sqrt{\overline{v^{2}}}}e^{-{2\pi V^{2}\over n\overline{\gamma^{2}}\ln
N}}\quad (V\lesssim V_{crit}(N)) 
\label{ATVsmall}
\end{equation}
\begin{equation}
T(V) ={\sqrt{\overline{\gamma^{2}}}\over 2\pi \sqrt{\overline{v^{2}}}V}\quad
(V\gtrsim V_{crit}(N))
\label{ATVlarge}
\end{equation}
The mean duration of the fluctuations is
\begin{equation}
\langle T\rangle ={2\over 3\sqrt{6}}{1\over\sqrt{n\overline{v^{2}}}}
\label{TaverageAppendix}
\end{equation}
and the diffusion coefficient
\begin{equation}
D={1\over 72}\biggl ({6\over \pi}\biggr
)^{1/2}\sqrt{\overline{\gamma^{2}}}\sqrt{\ln N}
\label{DAppendix}
\end{equation}
For a Gaussian distribution of the velocities, we recover the results of section
\ref{sec_statacceleration} appropriately generalized to account for a
distribution over the circulations. 

For ${\mb\rho}={\bf 0}$, the function $C({\mb \rho},{\mb \sigma})$ reduces to
\begin{equation}
C({\mb \sigma})=\int_{\gamma=-\infty}^{+\infty}  \int_{|{\bf v}|=0}^{+\infty}
\int_{0}^{+\infty}\tau(\gamma)\tau({\bf v})\biggl (1-J_{0}\biggl
({2\pi\sigma\over|\gamma|}v\Phi^{2}    \biggr )  \biggr ) {\gamma^{2}\over 2\pi
\Phi^{3}} d\gamma d^{2}{\bf v}d{\Phi}
\label{BC5}
\end{equation}  
Integrating by parts and using the identity
\begin{equation}
\int_{0}^{+\infty}{J_{1}(x)\over x}dx=1
\label{Bid}
\end{equation} 
we obtain
\begin{equation}
C({\mb \sigma})={1\over 2}\overline{|\gamma|}\overline{v}\sigma
\label{aBC6}
\end{equation} 
where 
\begin{equation}
\overline{v}=\int_{0}^{+\infty} \tau(|{\bf v}|)v 2\pi v dv
\label{Bvbar}
\end{equation}
is the average velocity of the vortices. Equation (\ref{AA8}) is changed to
\begin{equation}
A({\mb \sigma})=e^{-{n\overline{|\gamma|}\over 2}\overline{v}\sigma}
\label{BA6}
\end{equation} 
and the Cauchy distribution (\ref{WA3}) to
\begin{equation}
W({\bf A})={2\over \pi n^{2}\overline{|\gamma|}^{2}\overline{v}^{2}}{1\over
\biggl (1+{4 A^{2}\over
n^{2}\overline{|\gamma|}^{2}\overline{v}^{2}}\biggr)^{3/2}}
\label{BWA}
\end{equation} 

We find also
\begin{equation}
\langle V^{2}\rangle_{\bf A}={n\overline{\gamma^{2}}\over 4\pi}\ln N \quad
(A\lesssim A_{crit}(N))
\label{BV2A1}
\end{equation}
\begin{equation}
\langle V^{2}\rangle_{\bf A}={\overline{\gamma^{2}} A\over
2\pi\overline{|\gamma|} \overline{v}} \quad (A\gtrsim A_{crit}(N))
\label{BV2A2}
\end{equation}
with
\begin{equation}
A_{crit}(N)\sim{1\over 2}\overline{|\gamma |} n \overline{v}\ln N
\label{BAcrit}
\end{equation}
For a Gaussian distribution of the velocities, we recover the results of section
\ref{sec_statacceleration} appropriately generalized to account for a
distribution over the circulations.

\section{The case of a non neutral system}
\label{sec_nonneutral}

In this appendix, we consider the case where all vortices have the same
circulation $\gamma$ and we allow properly for the solid rotation of the system
that this distribution involves. We shall be particularly concerned with the
expression of the bivariate distribution $W({\bf V},{\bf A})$ at a point ${\bf
a}$ rotating around the center of the domain with velocity $\langle {\bf
V}\rangle ({\bf a}) ={1\over 2} n\gamma {\bf a}_{\perp}$. 

It is clear that equations (\ref{WW1}) (\ref{AA1}) (\ref{aPhi}) and (\ref{apsi})
remain valid in this more general situation if we make the substitutions ${\bf
r}\rightarrow {\mb\xi}={\bf r}-{\bf a}$ (relative distance) and ${\bf
v}\rightarrow {\bf v}'={\bf v}-\langle {\bf V}\rangle ({\bf a})$ (relative
velocity). According to equation  (\ref{Wgn}) that we shall prove ulteriorly,
the probability that a point vortex be found in ${\bf r}$ with velocity ${\bf
v}$ is:
\begin{equation}
\tau({\bf r},{\bf v})={1\over \pi R^{2}}\times {4\over n\gamma^{2}\ln
N}e^{-{4\pi\over n\gamma^{2}\ln N}({\bf v}-\langle {\bf V}\rangle ({\bf
r}))^{2}}
\label{txiw}
\end{equation}
We find it convenient to introduce the variable 
\begin{equation}
{\bf w}={\bf v}-\langle {\bf V}\rangle ({\bf r})={\bf v}'+\langle {\bf V}\rangle
({\bf a})-\langle {\bf V}\rangle ({\bf r}) ={\bf v}'-{1\over 2}n\gamma
{\mb\xi}_{\perp}
\label{w}
\end{equation}
In terms of this new variable, the acceleration (\ref{apsi}) becomes
\begin{equation}
{\mb\psi}({\mb\xi},{\bf v}')={\mb\psi}({\mb\xi},{\bf w})+{1\over 2}n\gamma
{\mb\Phi}_{\perp}
\label{psitrans}
\end{equation}
We can restore the initial form of $A({\mb\rho},{\mb\sigma})$ by defining
${\mb\rho}'={\mb\rho}-{1\over 2}n\gamma {\mb\sigma}_{\perp}$ in (\ref{AA1}).
Taking ${\mb\rho}'$ as a variable of integration in (\ref{WW1}), we find  that
the original system (\ref{WW1}) (\ref{AA1}) (\ref{aPhi}) (\ref{apsi}) (\ref{df})
is preserved if we make the substitutions ${\bf r}\rightarrow {\mb\xi}$, ${\bf
v}\rightarrow {\bf w}$ and  ${\bf A}\rightarrow {\bf A}-{1\over 2}n\gamma {\bf
V}_{\perp}$ (in the following, the dummy variable ${\mb\rho}'$ is renamed
${\mb\rho}$).

The characteristic function $A({\mb\rho},{\mb\sigma})$ can be written   
\begin{equation}
A({\mb\rho},{\mb\sigma})=e^{-n C_{a}({\mb\rho},{\mb\sigma})}
\label{ACapp}
\end{equation}
with
\begin{equation}
C_{a}({\mb\rho},{\mb\sigma})=\int \tau({\bf w})\biggl
(1-e^{i({\mb\rho}{\mb\Phi}+{\mb\sigma}{\mb\psi})}\biggr )d^{2}{\mb\xi}d^{2}{\bf
w}
\label{Capp}
\end{equation}
Since we are not at the center of the domain, the imaginary part of
$C_{a}({\mb\rho},{\mb\sigma})$ does not vanish. We have:
\begin{equation}
C_{a}({\mb\rho},{\mb\sigma})= C({\mb\rho},{\mb\sigma})-i
C'({\mb\rho},{\mb\sigma})
\label{Capp12}
\end{equation}
with
\begin{equation}
C({\mb\rho},{\mb\sigma})=\int \tau({\bf w})
(1-\cos({\mb\rho}{\mb\Phi}+{\mb\sigma}{\mb\psi}))d^{2}{\mb\xi}d^{2}{\bf w}
\label{Capp1}
\end{equation}
\begin{equation}
C'({\mb\rho},{\mb\sigma})=\int \tau({\bf w})
\sin({\mb\rho}{\mb\Phi}+{\mb\sigma}{\mb\psi})d^{2}{\mb\xi}d^{2}{\bf w}
\label{Capp2}
\end{equation}
The integrals $C({\mb\rho},{\mb\sigma})$ and $C'({\mb\rho},{\mb\sigma})$ have a
very different structure. The first integral  is dominated by relatively
proximate vortices and can be approximated with sufficient accuracy by the
function evaluated in section \ref{sec_statacceleration}. In the second
integral, the contribution of proximate vortices  vanishes by symmetry. As a
result, the integral is dominated by large values of $|{\mb\xi}|$ or,
equivalently, small values of $|{\mb\Phi}|$ and $|{\mb\psi}|$ . We can therefore
make the approximation $\sin( {\mb\rho}{\mb\Phi}+{\mb\sigma}{\mb\psi})\simeq
{\mb\rho}{\mb\Phi}+{\mb\sigma}{\mb\psi}$ and write 
\begin{equation}
C_{a}({\mb\rho},{\mb\sigma})=C({\mb\rho},{\mb\sigma})-i{\mb\rho}\int {\mb\Phi}
d^{2}{\mb\xi}-i{\mb\sigma}\int \tau({\bf w}) {\mb\psi}d^{2} {\mb\xi}d^{2}{\bf w}
\label{aCC3}
\end{equation}
The last integral cancels out due to the isotropy of the velocity distribution
$\tau({\bf w})$.  In the second integral, we recognize the mean-field velocity
created in ${\bf a}$ by the average distribution $n$ of the vortices:
\begin{equation}
\langle {\bf V}\rangle ({\bf a})=n\int {\mb\Phi}d^{2}{\mb\xi}={1\over 2}n\gamma
{\bf a}_{\perp}
\label{Vaverage}
\end{equation}
Therefore:
\begin{equation}
A({\mb\rho},{\mb\sigma})=e^{-n C({\mb\rho},{\mb\sigma})+i{\mb\rho}\langle {\bf
V}\rangle({\bf a})}
\label{CA1}
\end{equation}
Combining our previous results, we obtain
\begin{equation}
W({\bf V},{\bf A})={1\over 16\pi^{4}}\int e^{-i({\mb\rho}{\mb {\cal
V}}+{\mb\sigma}{\mb {\cal A}})} A({\mb\rho},{\mb\sigma})   d^{2}{\mb
\rho}d^{2}{\mb \sigma}
\label{CWV1}
\end{equation}
where $A({\mb\rho},{\mb\sigma})$ is the function considered in section
\ref{sec_statacceleration}. The bivariate distribution (\ref{CWV1}) at ${\bf
a}\neq {\bf 0}$ differs only from the distribution at the center of the domain
by replacing the velocity ${\bf V}$ by the fluctuating velocity ${\mb {\cal
V}}={\bf V}-\langle {\bf V}\rangle ({\bf a})={\bf V}-{1\over 2}n\gamma {\bf
a}_{\perp}$ and the acceleration ${\bf A}$ by the fluctuating acceleration ${\mb
{\cal A}}={\bf A}-{1\over 2}n\gamma {\bf V}_{\perp}$. We can account for a
spectrum of circulations among the vortices simply by changing $\gamma$ in
$\overline{\gamma}$ in the new terms and using the results of Appendix
\ref{sec_generalization}.

The mean life of a velocity fluctuations ${\mb {\cal V}}$ can be defined by 
\begin{equation}
T({\cal V})={|{\mb {\cal V}}|\over \sqrt{\langle {\cal A}^{2}\rangle_{{\mb
{\cal V}}}}}
\label{Tf}
\end{equation}
and the diffusion coefficient by
\begin{equation}
D={1\over 4}\int T({\cal V})W({\mb{\cal{V}}}){\cal V}^{2}d^{2}{\mb{\cal{V}}}
\label{Diffnonneutre}
\end{equation}
They have the the same functional form as (\ref{TV1})(\ref{TV2}) and
(\ref{Dexp}). Therefore, the results are unchanged in the case of a solid
rotation. For a differential rotation, the fluctuation time is related to the
local shear $\Sigma$ as investigated by\cite{chav98a} Chavanis (1998a) using an
approximation in which the point vortices are simply transported by the mean
flow. Therefore, the present theory gives the value of the fluctuation time in
regions where the shear cancels out. Of course, a more general calculation
should take into account the effect of both  the shear and  the dispersion of
particles, but this will not be considered in that paper. Note also that when
the system is inhomogeneous, the point vortices are subjected to a {\it
systematic drift} (\cite{chav98a} Chavanis 1998a) in addition to their diffusive
motion. This drift can be understood as the result of a polarization process.

\section{The case of vortex blobs}
\label{sec_blobs}

In reality, the vortices have a finite radius $a$ which is not necessarily
small. This finiteness is responsible for a maximum allowable velocity
$V_{max}\sim {\gamma\over 2\pi a}$ achieved when two vortices are at distance
$\sim 2a$ from each other. Higher velocities are forbidden because the subsequent
evolution is marked by merging events (see, e.g., Sire \& Chavanis 1999). It is
interesting to consider the distribution of velocity ${\bf V}$ and  acceleration
${\bf A}$ produced by a collection of vortex ``blobs'' with finite radius. This
problem was previously treated by \cite{jimenez} Jim\'enez (1996) and \cite{min}
Min {\it et al.} (1996) using numerical methods. The theory developed in this
article makes possible to obtain new analytical results.

Introducing a cut-off at $r=a$, equation (\ref{C6}) is replaced by
\begin{equation}
C_{N}({\mb \rho})={\gamma^{2}\rho^{2}\over 2\pi}\int_{\gamma\rho\over 2 \pi
R}^{\gamma\rho\over 2\pi a} (1-J_{0}(x)){dx\over x^{3}}
\label{D1}
\end{equation}
In the limit $N,R\rightarrow \infty$ with $n={N\over \pi R^{2}}$ finite, we
obtain
\begin{equation}
C({\mb \rho})={\gamma^{2}\rho^{2}\over 8\pi}\ln \biggl ({R\over a}\biggr )
\label{D2}
\end{equation} 
In the case of vortex blobs, the singularity at ${\mb\rho}={\bf 0}$ is removed
and the characteristic function $C_{N}({\mb \rho})$ is exactly quadratic.
Therefore $W({\bf V})$ is the Gaussian (\ref{Wgauss}) for all  $V\le V_{max}$.
There is no algebraic tail in the limit considered. However, the convergence to
the limit  distribution (\ref{Wgauss}) is still slow (see \cite{jimenez}
Jim\'enez 1996) and, in practice, the velocity distribution can differ
substantially from the Gaussian even for large values of $N$ (note that the
formalism presented in this paper could be used to study the dependence of the
velocity distribution with the number $N$ of vortices). 

More interesting is the distribution of the acceleration ${\bf A}$. In the case
of vortex blobs, equation (\ref{CC3}) is replaced by
\begin{equation}
C({\mb\sigma})={\gamma^{2}\over 2\pi}\int_{0}^{\gamma\over 2\pi a} (1-
e^{-{1\over 4}\pi n\ln N\sigma^{2}\Phi^{4}} ){d\Phi\over \Phi^{3}}  
\label{D3}
\end{equation}
After integrating by parts, one obtains:
\begin{equation}
C({\mb\sigma})=\pi a^{2} \biggl (e^{-{n\gamma^{4}\ln N\over 64\pi^{3}
a^{4}}\sigma^{2}}-1\biggr )+{\gamma^{2}\over 8}\sqrt{n\ln N} {\rm Erf} \biggl
({\gamma^{2}\over 8\pi^{2}a^{2}}\sqrt{n\pi\ln N}\sigma\biggr )\sigma
\label{D4}
\end{equation}
where 
\begin{equation}
 {\rm Erf}(x)={2\over\sqrt{\pi}}\int_{0}^{x}e^{-y^{2}}dy
\label{D5}
\end{equation}
is the ``Error function''.

For $\sigma\rightarrow 0$, the function $C({\mb\sigma})$ is quadratic in
$\sigma$: 
\begin{equation}
C({\mb\sigma})\sim {n\gamma^{4}\ln N\sigma^{2}\over 64\pi^{2} a^{2}}\qquad
(|{\mb\sigma}|\rightarrow 0)
\label{D6}
\end{equation}
implying that the distribution $W({\bf A})$ is Gaussian for large values of the
acceleration:
\begin{equation}
W({\bf A})\sim {16\pi a^{2}\over n^{2}\gamma^{4}\ln N}e^{-{16\pi^{2} a^{2}\over
n^{2}\gamma^{4}\ln N}A^{2}} \qquad  (|{\bf A}|\rightarrow +\infty)
\label{D7}
\end{equation}
Its variance is:
\begin{equation}
\langle A^{2}\rangle={n^{2}\gamma^{4}\over 16\pi^{2}a^{2}}\ln N
\label{D7b}
\end{equation}
For $\sigma\rightarrow +\infty$, the function $C({\mb\sigma})$ is linear in
$\sigma$: 
\begin{equation}
C({\mb\sigma})\sim{\gamma^{2}\over 8}\sqrt{n\ln N}\sigma\qquad (|{\mb
\sigma}|\rightarrow +\infty)
\label{D8}
\end{equation}
and we recover the Cauchy distribution (\ref{WA3}) for small values of $|{\bf
A}|$. Therefore, the distribution $W({\bf A})$  makes a smooth transition from
Cauchy (concave in a semi-log plot) for small fluctuations to  Gaussian  (convex
in a semi-log plot) for large fluctuations. It is likely that in between the
distribution passes through an {\it exponential tail} as observed  numerically
by \cite{min} Min {\it et al.} (1996). Of course, when $a$ is reduced, the
transition between the two regimes happens at larger fluctuations (see equation
(\ref{D7b})). According to equations (\ref{AP}) and (\ref{D7b}) the relevant
non-dimensional parameter to consider is the {\it area fraction} $n a^{2}$. In
decaying turbulence, the influence of an extended core should be manifest at the
beginning of the evolution.


\bigskip 

\begin{thebibliography}{99}

\bibitem{benzi}  {\small R. Benzi, M. Colella, M. Briscolini \& P. Santangelo, 
Phys.  Fluids A {\bf 4} (1992) 1036, ``A simple point vortex model for
two-dimensional decaying turbulence''.} 

\bibitem{bouchaud}  {\small J.P. Bouchaud \& A. Georges, Physics Reports  {\bf
195}, Nos 4 \& 5 (1990) 127-293, ``Anomalous diffusion in  disordered media:
statistical mechanisms, models and physical applications''.}

\bibitem{carnevale}  {\small G.F. Carnevale, J.C. McWilliams, Y. Pomeau, J.B.
Weiss \& W.R. Young, Phys. Rev. Lett.  {\bf 66} (1991) 2735, ``Evolution of
vortex statistics in two-dimensional turbulence''.}

\bibitem{chav96}  {\small P.H. Chavanis, Contribution \`a la m\'ecanique
statistique des tourbillons bidimensionnels. Analogie avec la relaxation
violente des syst\`emes stellaires. PhD thesis. Ecole Normale Sup\'erieure de
Lyon (1996).}

\bibitem{chav98a}  {\small P.H. Chavanis, Phys. Rev. E, {\bf 58} (1998a)
R1199-R1202, ``Systematic drift experienced by a point vortex in two-dimensional
turbulence''.}

\bibitem{chav98b}  {\small P.H. Chavanis, Annals of the New York Academy of
Sciences {\bf 867} (1998b) 120-141, ``From Jupiter's great red spot to the
structure of galaxies: statistical mechanics of two-dimensional vortices and
stellar systems'' in nonlinear dynamics and chaos in astrophysics: a festschrift
in honor of George Contopoulos.}

\bibitem{csr96}  {\small P.H. Chavanis, J. Sommeria \& R. Robert, Astrophy.
Journal, {\bf 471} (1996) 385-399, ``Statistical mechanics of two-dimensional
vortices and collisionless stellar systems''.}

\bibitem{chandr}  {\small S. Chandrasekhar, Astrophys. Journal {\bf 94} (1941)
511 ``A statistical theory of stellar encounters''.}

\bibitem{cn42}  {\small S. Chandrasekhar \& J. von Neumann, Astrophys. Journal
{\bf 95} (1942) 489, ``The Statistics of the gravitational field arising from a
random distribution of stars I. The speed of fluctuations''.}

\bibitem{cn43}  {\small S. Chandrasekhar \& J. von Neumann, Astrophys. Journal
{\bf 97} (1943) 1, ``The Statistics of the gravitational field arising from a
random distribution of stars II. The speed of fluctuations; dynamical friction;
spatial correlations''.}

\bibitem{hansen}  {\small A.E. Hansen, D. Marteau \& P. Tabeling, Phys. Rev. E
{\bf 58} (1998)  7261,  ``Two-dimensional turbulence and dispersion in a freely
decaying system''.}

\bibitem{holtsmark}  {\small J. Holtsmark, Ann. d. Phys. {\bf 58} (1919) 577.}

\bibitem{jimenez}  {\small J. Jim\'enez, J. Fluid Mech. {\bf 313} (1996)  223,
``Algebraic probability density tails in decaying isotropic two-dimensional
turbulence''.}

\bibitem{kandrup}  {\small H.E. Kandrup, Physics Report {\bf 63} (1980)  1,
``Stochastic gravitational fluctuations in a self-consistent mean field
theory''.}

\bibitem{min}  {\small I.A. Min, I. Mezic \& A. Leonard,  Phys.  Fluids {\bf 8}
(1996) 1169, ``L\'evy stable distributions for velocity and velocity difference
in systems of vortex elements''.}

\bibitem{novikov}  {\small E.A. Novikov, Sov. Phys. JETP {\bf 41} (1975) 937,
``Dynamics and statistics of a system of vortices''.}

\bibitem{sire}  {\small C. Sire \& P.H. Chavanis (1999), submitted to  Phys.
Rev. E (cond-mat/9912222), ``Numerical renormalization group of vortex
aggregation in 2D decaying turbulence: the role of three-body interactions''.}

\bibitem{smoluchowski}  {\small M. Smoluchowski, Phys. Zs. {\bf 17} (1916) 557.}

\bibitem{weissmcwilliams}  {\small J.B. Weiss \& J.C. McWilliams,  Phys.
Fluids A {\bf 5} (1993) 608,  ``Temporal scaling behavior of decaying
two-dimensional turbulence''.}

\bibitem{weiss}  {\small J.B. Weiss, A. Provenzale \& J.C. McWilliams,  Phys.
Fluids {\bf 10} (1998) 1929,  ``Lagrangian Dynamics in High-Dimensional
Point-vortex Systems''.}



\end{thebibliography}
\end{document}